\documentclass[12pt]{iopart}

\usepackage{iopams} 

\usepackage{graphicx}

\usepackage{color}

\newcommand{\be}{\begin{equation}}
\newcommand{\ee}{\end{equation}} 

\def\nn{\nonumber} 
\begin{document}
\title[Long-wavelength instabilities in a system  of interacting active particles]
      {Long-wavelength instabilities in a system  of interacting active particles}


\author{Zahra Fazli}

\address{Department of Physics, Institute for Advanced Studies in Basic Sciences (IASBS), Zanjan 45137-66731, Iran}

\author{Ali Najafi}
\address{Department of Physics, Institute for Advanced Studies in Basic Sciences (IASBS), Zanjan 45137-66731, Iran}

\address{Research Center for Basic Sciences \& Modern Technologies (RBST),  Institute for Advanced Studies in Basic Sciences (IASBS), Zanjan 45137-66731, Iran
}
\ead{najafi@iasbs.ac.ir}
\vspace{12pt}

\begin{abstract}
Based on a microscopic model, we develop a continuum description for a suspension of  microscopic 
self propelled particles. With this continuum description we study the role of long-range interactions in destabilizing  macroscopic ordered phases that are developed  by short-range interactions. 
Long-wavelength fluctuations can destabilize both  isotropic  and also symmetry broken polar phase 
in a suspension of dipolar particles. The instabilities  in 
a suspension of pullers (pushers) arise from  splay (bend) fluctuations.  Such instabilities are not seen in a suspension of 
quadrupolar particles.
\end{abstract}

\vspace{2pc}
\maketitle
%
%

\section{Introduction}
Dynamics of a suspension of interacting active particles, as a  non-equilibrium problem in statistical mechanics,  
has attracted  enormous interests in recent years \cite{Ramaswamy,Marchetti3,Vicsek,Ramaswamy5}. 
Systems like schools of fishes and birds \cite{Becco,Cavagna,Nagy}, bacterial colonies \cite{Peruani,Tokita,Lushi,Sokolov}, gels of bio-polymers \cite{Prost} and
interacting active Janus particles \cite{Theurkauff,Speck,Bayati} show a wide range of fascinating physical behavior.  
Coherent collective motions, long-range orientational order, large number fluctuations and pattern formations  are examples of such phenomena \cite{Peruani,Narayan,Schaller}.

Existence of  long-range order in two-dimensional active systems \cite{Schaller,Sumino} 
seems to be in contrast with Mermin-Wagner theorem at first glance.  As a result of 
that theorem,  true order in low dimensional equilibrium systems is not possible \cite{Mermin}, 
but a theoretical work based on renormalization group analysis 
by Toner and Tu has revealed a physical scenario in which,   non-equilibrium nature of  
active systems  can provide conditions for true order in lower dimensions \cite{Tu}.   
Usually, short-range interactions are responsible for developing  ordered phases but the role of 
long-range interactions, needs to be considered carefully \cite{Marchetti2,Stark3,Shelley,Behmadi}. In a system 
composed of active particles suspended in 
aqueous   media, hydrodynamic interactions provide long-range forces that can propagate like $1/r^2$  to long distances.
In  systems with long-range interactions, macroscopic ordered phases developed by short-range 
interactions are under dynamical instabilities due to long-wavelength fluctuations \cite{Shelley2,inst1,inst2,inst3}. Such instabilities are 
very sensitive to  microscopic details of  swimming mechanisms  that 
can  distinguish   a pusher, puller 
or a neutral swimmer \cite{Rao1,Stark}. Studying such instabilities is the main purpose of current article.

In addition to   numerical studies \cite{Peruani2,Vicsek2,Peruani3}, continuum descriptions can provide  analytical tools in dealing with 
such non-equilibrium  systems. Microscopic derivations \cite{Marchetti5} and symmetry arguments \cite{Tu} are two approaches  
that can provide the governing equations for macroscopic continuum fields. While a large amount  of works are devoted 
to the symmetry based theories \cite{Tu1,Ramaswamy4,Mishra},  less efforts are concentrated  on microscopic derivations \cite{Marchetti2,Bertin}.

In this article we  
aim to use a microscopic approach and obtain the equations of macroscopic description.  
 The continuum description derived from a 
microscopic model in this article, will allow us to study the role of long-range interactions in  instabilities observed in 
active suspensions. 
Theories based on symmetry 
arguments reveals qualitative features of the long-wavelength instabilities in active suspensions.
Microscopic based models can help us to understand the origin of instabilities  more    
quantitatively.    We will show that both isotropic and polar phases that can appear in active systems are unstable 
with respect to long-wavelength fluctuations.

The structure of this article is as follows: In section \ref{sec2}, we present  the hydrodynamic 
details of our microscopic model and introduce  long- and short-range interactions between swimmers.  Then in 
section \ref{sec3}, we describe the dynamics of a suspension of many swimmers in terms of Langevin 
and Smoluchowski descriptions. Furthermore, in this section, we simplify the description by considering 
mean field approximation.  
In section \ref{sec4}, we derive a continuum description for the system. 
Dynamical equations, their steady state solutions and instability analysis  are presented in this section. 
Finally, discussion and summary are presented in section \ref{sec5}.


\section{Hydrodynamic model for   micro-swimmers}
\label{sec2}
We start with a  microscopic model for a minimal autonomous micro-swimmer  that can propel itself 
at aqueous  media.  Theoretical arguments based on symmetry grounds show that a minimum number of two internal degrees of freedom is necessary to capture the hydrodynamic details of a micro-swimmer \cite{Purcell}.  To construct the model swimmer, consider three spheres with 
radii  $a$, connected linearly by two arms with variable lengths given by $L^f$ and $L^b$. 
We label the spheres by $f$ (front), $b$ (back) and  $m$ (middle).
It is shown that harmonic changes in the arm lengths with a phase lag between arms, will result a non zero 
swimming velocity for this system \cite{3SJPC,3Sfaez}.  
To see how the above swimmer can work, one needs to solve the hydrodynamic equations for the 
ambient fluid that are coupled to the motion of spheres.  At the scale of micrometer with velocity range 
about micrometer per second in water, the linear Stokes equation governs the dynamics of  the fluid.   
Assuming that the arms are thin enough to neglect their hydrodynamic effects and eliminating the fluid degrees of freedom, one can reach to effective equations that govern the dynamics of spheres alone. 
Such equations are linear relations between the velocity of spheres and hydrodynamic forces acting by spheres 
on the fluid \cite{Pozrikidis}:  
\begin{eqnarray}\label{vs}
v^{m}_i =\sum_{j,n} \,O^{mn}_{ij}\,f^{n}_j,
\end{eqnarray}
where $f^{n}_j$ ($v^{n}_j$) denotes the $j$-th component of the force (velocity) of 
sphere $n$ and the details of the hydrodynamic interactions are given by the kernel 
$O^{mn}_{ij}$.  This hydrodynamic kernel is a function of the size of spheres and their relative 
position. Denoting the distance between spheres $m$ and $n$ by ${\bf d}= \textbf{x}^m-\textbf{x}^n$ and fluid viscosity by $\eta$ and in the limit of $d\gg a$,  Oseen's tensor provides an 
approximation for the hydrodynamic kernel \cite{Pozrikidis}:
\begin{equation}
\label{cases}
O^{mn}_{ij}=\cases{
\frac{1}{8\pi\eta d}\left(\delta_{ij}+\hat{d}_i\hat{d}_j\right) &for $m\neq n$\\
\frac{\delta_{ij}}{6\pi\eta a}&for $m=n$\\}.
\end{equation}
As the swimmer is autonomous, one needs to add the conditions of zero total force and zero total torque to the above dynamical equations. The above  relations and the constraints  that prescribe the 
dynamics of arm lengths provide a complete set of dynamical equations that can 
fully determine the state of the swimmer, including its speed, direction and forces.  Velocities and forces 
averaged over time, are the quantities that we are interested to know.   
To express the results, let us assume that the arms oscillate  around a mean value $\ell$ as:
 \begin{eqnarray}
L^f(t)=\ell+u^f(t),\qquad L^b(t)=\ell(1+\delta)+u^b(t),
\end{eqnarray}
where $u^{f}$ and $u^{b}$ are periodic functions of time and 
$\delta$  is a parameter that makes the swimmer geometrically asymmetric. 
After solving the above equations, the average swimming velocities (linear and angular) and forces acting on the
fluid read as \cite{3Sfaez}:
\begin{eqnarray} \label{1}
\textbf{v}_0=v_0\,\hat{\textbf{t}},\qquad\qquad\boldsymbol\Omega_0=\textbf{0},
\end{eqnarray}
\numparts
\begin{equation}
\langle \textbf{f}^{\,f}\rangle=-\frac{5}{4}\pi\eta(\frac{a}{\ell})^2(1-\frac{17}{5}\delta)\Phi\,\hat{\textbf{t}},~~
\langle \textbf{f}^{\,b}\rangle=-\frac{5}{4}\pi\eta(\frac{a}{\ell})^2(1+\frac{7}{5}\delta)\Phi\,\hat{\textbf{t}},
\end{equation}
\endnumparts
where $v_0=-\frac{7}{12}(\frac{a}{\ell^2})(1-\delta)\Phi$ and ${\hat {\bf t}}$ represents the direction of the swimmer and  
$\Phi=\langle u^f\dot{u}^b\rangle$ with  $\langle\dots\rangle$ shows the averaging over time. 
Additionally $\langle \textbf{f}^{\,m}\rangle=-\langle \textbf{f}^{\,b}\rangle-\langle \textbf{f}^{\,f}\rangle$.  In writing 
the above results, we have assumed that $a\ll \ell$, $u^f\ll\ell$, $u^b\ll\ell$ and $\delta\ll 1$.  
Throughout this paper we choose $\Phi<0$, so that $v_0>0$. 

\begin{figure}
\centering
\includegraphics[width=10cm]{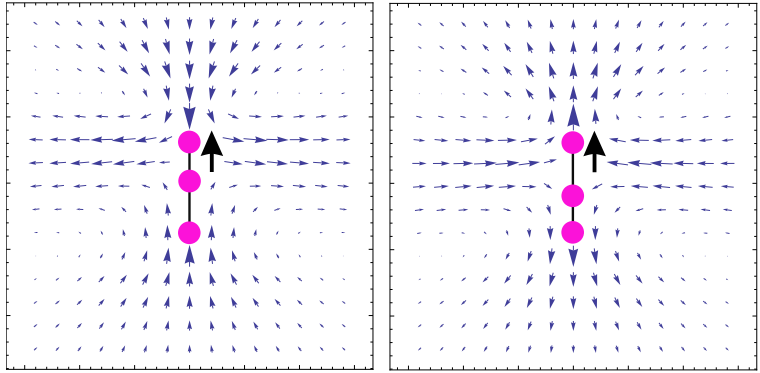}
\caption{Velocity field of a puller (left) and a pusher (right). In both cases, the swimmer moves upward, along the bolded arrow.  Velocity filed decreases as  $1/r^2$.}
\label{fig1}
\end{figure}

When considering the force distribution of a swimmer, the asymmetry parameter $\delta$ plays an essential  role. For a symmetric swimmer ($\delta=0$), the force distribution shows a quadrupolar 
field while for asymmetric case ($\delta\neq 0$) it shows a dipolar filed \cite{Moradi}. Defining the 
force dipole tensor as: $
\Gamma_{ij}=\sum_m x_{i}^{m}\,f^{\,m}_{\,j}$, 
we can calculate it as:
\begin{eqnarray}
\boldsymbol\Gamma=-\frac{29}{10}f_0\,\ell\,\delta\,\hat{\textbf{t}}\hat{\textbf{t}},
\end{eqnarray}
where $f_0=\frac{30}{7}\pi\eta \,a\,v_0$. 
Based on the observation that how the driving force of the motion is 
located at the head or at the tail of the swimmer, we can divide  the dipolar swimmers to two categories of 
pushers and pullers.
For pushers, the driving force comes from the tail 
while for pullers, the driving force comes from  the head. 
In asymmetric three-sphere swimmer with $\Phi<0$, $\delta>0$ corresponds to a puller ($\Gamma_{tt}<0$) 
and $\delta<0$ 
results a  pusher ($\Gamma_{tt}>0$). 
For a puller (pusher), back (front) arm of the swimmer is longer than the front (back) arm.  
Figure~\ref{fig1}, shows the flow filed pattern for both pusher and puller. There is a fundamental 
difference between the flow patterns for pushers and pullers. 
At the next parts we will see that the hydrodynamic interaction between the swimmers will crucially 
depend on the sign of $\delta$.  
\subsection{Long-range interactions}
Since we want to consider  a suspension of micro-swimmers, we need to calculate the  
hydrodynamic interactions between them. The above model of micro-swimmer 
allows us to obtain analytic 
formula for the interactions. The details of such calculations are similar to the case of a single swimmer and 
have been studied in details elsewhere \cite{Farzin,Yeomans4}. Here we only present the final results. 
Consider two swimmers located at positions ${\bf r}$ and ${\bf r}^\prime$ with orientations 
given by $\hat{\textbf{t}}$ and $\hat{\textbf{t}}^\prime$. Taking into account the hydrodynamic 
long-range interactions between the swimmers, the linear and angular velocities of the first 
swimmer averaged over the oscillations of both swimmers read as:
\begin{eqnarray}\label{T}
\textbf{V}^L\left(\textbf{r},{\bf r}',\hat{\textbf{t}},\hat{\textbf{t}}'\right)=a_1\,\left(\frac{\ell}{R}\right)^2\,\textbf{G}_1+\left(\frac{\ell}{R}\right)^3\,\left(a_2\,\textbf{G}_2+a_3\,\textbf{G}_3\right),
\end{eqnarray}
\begin{eqnarray}\label{R}
\boldsymbol\Omega^L\left(\textbf{r},{\bf r}',\hat{\textbf{t}},\hat{\textbf{t}}'\right)=a_4\,\left(\frac{\ell}{R}\right)^3\,\textbf{G}_4+\left(\frac{\ell}{R}\right)^4\,\left(a_5\,\textbf{G}_5+a_6\,\textbf{G}_6\right).
\end{eqnarray}
where $\textbf{R}= \textbf{r}-{\bf r}'$ and superscript $L$ denotes long-range interaction.  
The coefficients are given by:
\begin{eqnarray}
a_1=-\frac{87}{56}\left(\frac{a}{\ell}\right)\delta v_0,\qquad a_2=-\frac{6}{7}\left(2+\delta \right)v_0,\qquad a_3=-\frac{1}{2}a_2,\nn\\
a_4=-\frac{1}{\ell}a_1,\qquad a_5=-\frac{1}{\ell}a_2,\qquad a_6=\frac{3}{2}\left(\frac{1}{\ell}\right)\left(2-\delta \right)v_0.
\end{eqnarray}
Regarding the above results for interaction, the terms proportional to $a_1$ and $a_4$ represent the dipolar contributions and the other terms show the quadrupolar 
contributions. 
Vectors $\textbf{G}_1,\cdots,\textbf{G}_6$ are complex functions of relative displacement and orientation of the swimmers and are given by:
\begin{eqnarray}
\textbf{G}_1=-3M_{ij}(\hat{\textbf{R}})\,\hat{t_i}^\prime\hat{t_j}^\prime\hat{\textbf{R}},
\end{eqnarray}
\begin{eqnarray}
\textbf{G}_2=\frac{3}{2}M_{ij}(\hat{\textbf{R}})\,\hat{t_i}'\hat{t_j}'\hat{\textbf{t}}'+\frac{3}{2}M_{ijk}(\hat{\textbf{R}})\,\hat{t_i}'\hat{t_j}'\hat{t_k}'\hat{\textbf{R}},
\end{eqnarray}
\begin{eqnarray}
\textbf{G}_3=-3M_{ij}(\hat{\textbf{R}})\,\hat{t_i}'\hat{t_j}'\hat{\textbf{t}}+3M_{ijk}(\hat{\textbf{R}})\,\hat{t_i}'\hat{t_j}'\hat{t_k}\hat{\textbf{R}},
\end{eqnarray}
\begin{eqnarray}
\textbf{G}_4=3M_{ijk}(\hat{\textbf{R}})\,\hat{t_i}'\hat{t_j}'\hat{t_k}\hat{\textbf{R}},
\end{eqnarray}
\begin{eqnarray}
\textbf{G}_5=\frac{3}{2}M_{ijk}(\hat{\textbf{R}})\,\hat{t_i}'\hat{t_j}'\hat{t_k}\hat{\textbf{t}}'-\frac{15}{2}M_{ijkl}(\hat{\textbf{R}})\,\hat{t_i}'\hat{t_j}'\hat{t_k}'\hat{t}_{l}\hat{\textbf{R}},
\end{eqnarray}
\begin{eqnarray}
\textbf{G}_6=-\frac{15}{2}M_{ijkl}(\hat{\textbf{R}})\,\hat{t_i}'\hat{t_j}'\hat{t_k}\hat{t_l}\hat{\textbf{R}},
\end{eqnarray}
where summation over repeated indices is assumed and:
\begin{eqnarray}
&M_{ij}(\hat{\textbf{R}})=\hat{R}_i\hat{R}_j-\frac{1}{3}\,\delta_{ij},\qquad M_{ijk}(\hat{\textbf{R}})=-R^4\partial_k\left(\frac{M_{ij}}{R^3}\right),\nn\\
&M_{ijkl}(\hat{\textbf{R}})=-\frac{R^5}{5}\partial_l\left(\frac{M_{ijk}}{R^4}\right),
\end{eqnarray}
where we have used the short hand notation: $\partial_i={\partial}/{\partial R_i}$.
To obtain  the above hydrodynamic interactions we have assumed that the swimmers are very far, $R\gg\ell$, and we have also averaged over the internal motion of the swimmers.
As it is seen from equations (\ref{T}) and (\ref{R}), the first non-zero terms in the hydrodynamic 
interaction, the terms that are proportional to $(\frac{\ell}{R})^2$ in linear velocity and 
$(\frac{\ell}{R})^3$ in rotational velocity, are  
proportional to $\delta$. This is the contribution from dipolar filed of the asymmetric swimmers. Such contribution  changes sign for 
pushers and pullers \cite{Lauga,Elgeti,Drescher}.

Rich dynamical behavior that includes coherent motion in two interacting swimmers suggests to 
see interesting phases in a system with many interacting swimmers \cite{Yeomans4,coherentcoupling}.
At next sections we will see how thermodynamic behavior of a suspension of micro-swimmers depends 
on the nature of two particle interactions.  

\subsection{Short-range interactions}
\label{SRI}
As one can see from equations (\ref{T}) and (\ref{R}), the long-range hydrodynamic interactions that 
we have obtained are valid only at large distances, they diverge
at short distances. Due to the complexity of hydrodynamics  at short distances, it is not possible to obtain simple 
analytic results for   short-range part of the interactions. 
We can use  an approximate phenomenological model that takes into account the short-range part of the interactions. 
\begin{figure}[htbp]
\centering
\includegraphics[width=8cm]{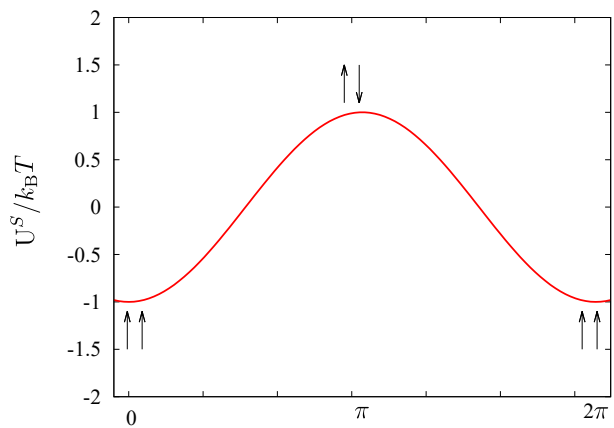}
\caption{
Short-range  alignment interaction, ${\rm U}^S$, between the swimmers is plotted as a function of the angle between their orientations. The potential has a minimum when  two swimmers are aligned. Here we assumed $U_0=1$.
}
\label{fig2}
\end{figure}
Inspired by the well known Vicseck's model \cite{Vicsek2}, we consider a short-range interaction potential as:
\begin{equation}
\label{cases}
{{\rm U}^{S}(\textbf{r},\textbf{r}^\prime,\hat{{\bf t}},\hat{{\bf t}}^\prime
)}=\cases{
  -{k_{\rm B}T}U_0 \,\hat{\textbf{t}}\cdot\hat{\textbf{t}}^\prime \qquad & for $
  R \le \ell_c,$
\\
 0 \qquad\qquad\qquad &for $R > \ell_c.$\\}
\end{equation}
where $\ell_c$ is a crossover 
length-scale that separates short- and long-range interactions. 
We will assume that long-range hydrodynamic interactions act only for  swimmers having distance 
larger than $\ell_c$.  As seen from figure~\ref{fig2}, the above potential tends to align nearby particles. 
The crossover length $\ell_c$ has the same order of magnitude  as the linear dimension of swimmers given by $2\ell$.
It should be mentioned that the above  interaction  does not  consider all informations  of the short-range  interactions in real systems, this model  only takes into account the steric interaction between the nearby swimmers. 
In terms of the potential energy, the short-range velocities  can be written as: 
\begin{eqnarray}
\textbf{V}^S(\textbf{r},\textbf{r}^\prime,\hat{{\bf t}},\hat{{\bf t}}^\prime)=-\frac{1}{k_{\rm B}T}\,\textbf{D}\cdot\boldsymbol\nabla {\rm U}^S,
\label{st}
\end{eqnarray}
\begin{eqnarray}
\boldsymbol\Omega^S(\textbf{r},\textbf{r}^\prime,\hat{{\bf t}},\hat{{\bf t}}^\prime)=-\frac{D_R}{k_{\rm B}T}\, \boldsymbol\nabla_R{\rm U}^S,
\label{sr}
\end{eqnarray}
where $\boldsymbol\nabla$ and $\boldsymbol\nabla_R\equiv\hat{\textbf{t}}\times \partial/\partial \hat{\textbf{t}}$ are translational and rotational gradient operators.

In above equations, $\textbf{D}$ is the translational diffusion tensor of a micro-swimmer and for a swimmer with 
elongated geometry we can decompose it to its  parallel, $D_\parallel$, and perpendicular, $D_\perp$, components:
\begin{eqnarray}
{\rm D}_{ij}=D_\parallel \hat{t}_i\hat{t}_j+D_\perp\left(\delta_{ij}-\hat{t}_i\hat{t}_j\right),
\end{eqnarray}
and $D_R$ is the rotational diffusion coefficient.
Through hydrodynamic calculations, we can calculate  the translational and rotational diffusion coefficients \cite{Kim,Happel}. 

Details of such calculations are presented in  \ref{app1}.  The final results are:
\begin{eqnarray}
D_\parallel=\frac{k_{\rm B}T}{18\pi\eta a}\Bigg[1+\frac{5}{2}\left(1-\frac{\delta}{2}\right)\left(\frac{a}{\ell}\right)+\mathcal{O}\left(\frac{a}{\ell}\right)^2\Bigg],
\end{eqnarray}
\begin{eqnarray}
D_\perp=\frac{k_{\rm B}T}{18\pi\eta a}\Bigg[1+\frac{5}{4}\left(1-\frac{\delta}{2}\right)\left(\frac{a}{\ell}\right)+\mathcal{O}\left(\frac{a}{\ell}\right)^2\Bigg],
\end{eqnarray}
\begin{eqnarray}
D_R=\frac{k_{\rm B}T}{12\pi\eta a\,\ell^2}\Bigg[(1-\delta)-\frac{3}{8}\left(1-\frac{3}{2}\delta\right)\left(\frac{a}{\ell}\right)+\mathcal{O}\left(\frac{a}{\ell}\right)^2\Bigg],
\end{eqnarray}

\section{Dynamics of a suspension}
\label{sec3}
\subsection{Langevin Dynamics}
Let us consider a dilute suspension of ${\cal N}$ micro-swimmers moving in a three-dimensional fluid medium with 
temperature $T$.  
To describe the dynamics of the suspension, we can start with  Langevin description for each 
micro-swimmer as: 
\begin{eqnarray}\label{t}
\partial_t \textbf{r}_\alpha=v_0\,\hat{\textbf{t}}_\alpha+\sum_{\beta\neq \alpha}\textbf{V}
^{int}\left(\textbf{r}_\alpha,\textbf{r}_\beta,\hat{\textbf{t}}_\alpha,\hat{\textbf{t}}_\beta\right)+\boldsymbol\eta_\alpha^{T}(t),
\end{eqnarray}
\begin{eqnarray}\label{r}
\partial_t\hat{\textbf{t}}_\alpha=\sum_{\beta\neq \alpha}\boldsymbol\Omega^{int}\left(\textbf{r}_\alpha,\textbf{r}_\beta,\hat{\textbf{t}}_\alpha,\hat{\textbf{t}}_\beta\right)\times \hat{\textbf{t}}_\alpha+\boldsymbol\eta_\alpha^{R}(t),
\end{eqnarray}
where $\textbf{r}_\alpha$ denotes  the position vector for the hydrodynamic center of $\alpha$'th 
swimmer ($\alpha=1,\dots,{\cal N}$) and $\hat{\textbf{t}}_\alpha$ for its director. 
Hydrodynamic center is defined in appendix. 
 In above relation the summation 
is over all other swimmers ($\beta\neq \alpha$).   
$\textbf{V}^{int}$ and $\boldsymbol\Omega^{int}$ are interaction contributions to the translational and rotational velocities of  swimmers. We consider two types of interactions between the swimmers: a short-range alignment interaction and a long-range one that is due to the fluid-mediated interactions between the swimmers. So $\textbf{V}^{int}$ and $\boldsymbol\Omega^{int}$ contain two terms:
\begin{eqnarray}
\textbf{V}^{int}\left(\textbf{r}_\alpha,\textbf{r}_\beta,\hat{\textbf{t}}_\alpha,\hat{\textbf{t}}_\beta\right)=\textbf{V}^L+\textbf{V}^S,
\end{eqnarray}
\begin{eqnarray}
\boldsymbol\Omega^{int}\left(\textbf{r}_\alpha,\textbf{r}_\beta,\hat{\textbf{t}}_\alpha,\hat{\textbf{t}}_\beta\right)=\boldsymbol\Omega^L+\boldsymbol\Omega^S,
\end{eqnarray}
where in the last section we have obtained the long- and short-range part of the interaction. $\boldsymbol\eta_\alpha^{T}(t)$ and $\boldsymbol\eta_\alpha^{R}(t)$ are stochastic terms due to the random forces which swimmer $\alpha$ receives from the molecules of the ambient fluid.
The random forces obey the statistics of a Gaussian noise as:
\begin{eqnarray}
\langle\eta_{\alpha,i}^T(t)\eta_{\beta,j}^T(t^{\prime})\rangle= {\rm D}_{ij}\,\delta_{\alpha\beta}\,\delta(t-t^{\prime}),
\end{eqnarray}
\begin{eqnarray}
\langle\eta_{\alpha,i}^R(t)\eta_{\beta,j}^R(t^{\prime})\rangle= D_{R}\,\delta_{ij}\,\delta_{\alpha\beta}\,\delta(t-t^{\prime}).
\end{eqnarray}

\subsection{Statistical  description}
\label{continuum}
In order to have a probabilistic description for a suspension composed of ${\cal N}$ particles, we denote the   
${\cal N}$-body probability distribution function by: $\Psi_{\cal N} (\textbf{r}_1,\hat{\textbf{t}}_1,\cdots,\textbf{r}_{\cal N},\hat{\textbf{t}}_{\cal N},t)$. The distribution
function  is the probability to find the $\alpha$'th swimmer at position ${\bf r}_{\alpha}$ with 
 orientation given by ${\hat {\bf t}}_\alpha$ at time $t$. This distribution function obeys the following 
  normalization condition:
\begin{eqnarray}
\prod_{\alpha=1}^{\cal N}\int {\rm d}\textbf{r}_\alpha {\rm d}\hat{\textbf{t}}_\alpha\,\Psi_{\cal N}=1,
\end{eqnarray}
and it satisfies the following continuity equation:
\begin{eqnarray}
\partial_t\Psi_{\cal N}=-\left(\sum_{\alpha=1}^{\cal N} \frac{\partial}{\partial \textbf{r}_\alpha}\right)\cdot\textbf{J}_{\cal N}^T-\left(\sum_{\alpha=1}^{\cal N} \hat{\textbf{t}}_\alpha\times\frac{\partial}{\partial\hat{\textbf{t}}_\alpha}\right)\cdot\textbf{J}_{\cal N}^R,
\end{eqnarray}
where $\textbf{J}_{\cal N}^T$ and $\textbf{J}_{{\cal N}}^{R}$ are translational and rotational ${\cal N}$-body fluxes. At very low volume fraction of swimmers where, the distance between the swimmers is  larger than the 
size of swimmers, we can treat the system in the mean field level. In this case the ${\cal N}$-body 
distribution function can be given in terms of  single particle distribution function as:
\begin{eqnarray}
\Psi_{\cal N} &=\psi \left(\textbf{r}_1,\hat{\textbf{t}}_1,t\right)\cdots\psi\left(\textbf{r}_{\cal N},\hat{\textbf{t}}_{\cal N},t\right).
\end{eqnarray}
Using this assumption, the single particle distribution function,  
$\psi\left(\textbf{r},\hat{\textbf{t}},t\right)$, obeys the following Smoluchowski equation: 
\begin{eqnarray}
\partial_t\psi=-\boldsymbol\nabla\cdot\textbf{J}^T-\boldsymbol\nabla_R\cdot\textbf{J}^R,
\label{spd}
\end{eqnarray}
where $\textbf{J}^T$ and $\textbf{J}^R$ are translational and rotational one-body fluxes and are given 
by:
\begin{eqnarray}
\textbf{J}_T=\left[v_0\hat{\textbf{t}}+\overline{\textbf{V}}^{int}\,\right]\psi-\textbf{D}\cdot\boldsymbol\nabla\psi,
\end{eqnarray}
\begin{eqnarray}
\textbf{J}_R=\overline{\boldsymbol\Omega}^{int}\,\psi-D_R\,\boldsymbol\nabla_R\,\psi.
\end{eqnarray}
Mean field translational and rotational velocities  are denoted by
 $\overline{\textbf{V}}^{int}$ and $\overline{\boldsymbol\Omega}^{int}$.  These mean field terms should be calculated by integrating over the positions and orientations of all   swimmers as:
\begin{eqnarray}
\overline{\textbf{V}}^{int}\left(\textbf{r},\hat{\textbf{t}},t\right)=\int \rmd\textbf{r}^\prime \rmd\hat{\textbf{t}}^\prime \,\textbf{V}^{int}(\textbf{r},\textbf{r}^\prime,\hat{\textbf{t}},\hat{\textbf{t}}^\prime)\,\psi(\textbf{r}^\prime,\hat{\textbf{t}}^\prime,t),
\label{mft}
\end{eqnarray}
\begin{eqnarray}
\overline{\boldsymbol\Omega}^{int}\left(\textbf{r},\hat{\textbf{t}},t\right)=\int \rmd\textbf{r}^\prime \rmd\hat{\textbf{t}}^\prime \,\boldsymbol\Omega^{int}(\textbf{r},\textbf{r}^\prime,\hat{\textbf{t}},\hat{\textbf{t}}^\prime)\,\psi(\textbf{r}^\prime,\hat{\textbf{t}}^\prime,t).
\label{mfr}
\end{eqnarray}
In order to study the dynamics of an active system composed of interacting particles, we 
proceed and consider the dynamics of   moments of distribution function. Density field 
$\rho({\bf r},t)$, polarization ${\bf P}({\bf r},t)$ and nematic order parameter ${\bf N}({\bf r},t)$ 
are the first three moments of the distribution function that are defined as:
\begin{eqnarray}
\rho(\textbf{r},t)=\int \rmd\hat{\textbf{t}}\,\psi(\textbf{r},\hat{\textbf{t}},t),
\end{eqnarray}
\begin{eqnarray}
\rho(\textbf{r},t)\textbf{P}(\textbf{r},t)=\int \rmd\hat{\textbf{t}}\,\hat{\textbf{t}}\,\psi(\textbf{r},\hat{\textbf{t}},t),
\end{eqnarray}
\begin{eqnarray}
\rho(\textbf{r},t)\textbf{N}(\textbf{r},t)=\int \rmd\hat{\textbf{t}}\,\left(\hat{\textbf{t}}\hat{\textbf{t}}-\frac{\textbf{I}}{3}\right)\psi(\textbf{r},\hat{\textbf{t}},t).
\end{eqnarray}
Using   equation (\ref{spd}), we can obtain  equations that govern the 
dynamics of above continuum fields.  As a result of such  equations we see that the dynamics of $n$'th moment is coupled  to 
the dynamics of $(n-1)$'th moment. So we need to cut the hierarchy at some point. As an approximation, we 
neglect the third (and higher) moment and  cut the equations at second moment. In this case and 
in terms of density, polarization and nematic order, the distribution function can be constructed as:   
\begin{eqnarray}
\psi(\textbf{r},\hat{\textbf{t}},t)=\rho(\textbf{r},t)\left(\frac{1}{4\pi}+\frac{3}{4\pi}\,\hat{\textbf{t}}\cdot\textbf{P}(\textbf{r},t)+\frac{15}{8\pi}\,\left(\hat{\textbf{t}}\hat{\textbf{t}}-\frac{\textbf{I}}{3}\right):\textbf{N}(\textbf{r},t)\right).
\end{eqnarray}
\subsection{Mean field interactions}
Before deriving  the dynamical equations for  continuum fields, we need to calculate the 
 mean field form of interaction terms. As discussed before, the interaction between swimmers 
 has two contributions, short-  and long-range  parts as: 
 \begin{eqnarray}
&&\overline{\textbf{V}}^{int}\left(\textbf{r},\hat{\textbf{t}},t\right)=\overline{\textbf{V}}^{S}+\overline{\textbf{V}}^{L}\nonumber\\
&&\overline{\boldsymbol\Omega}^{int}\left(\textbf{r},\hat{\textbf{t}},t\right)=\overline{\boldsymbol\Omega}^{S}+\overline{\boldsymbol\Omega}^{L}.
\end{eqnarray}
To obtain the short-range contribution we 
 need to insert the two-body interactions from equations (\ref{st}) and (\ref{sr}) into equations (\ref{mft}) and (\ref{mfr}) then, calculate the integrations. 
To obtain the final results, the following integral should be performed:
\begin{eqnarray}
\overline{{\rm U}}^S(\textbf{r},\hat{\textbf{t}},t)=\int {\rm d}\textbf{r}^\prime {\rm d}\hat{\textbf{t}}^\prime\,{\rm U}^{S}\,\psi(\textbf{r}^\prime,\hat{\textbf{t}}^\prime,t).
\end{eqnarray}
Now, as the interaction is short-range, we can expand  $\psi(\textbf{r}^\prime,\hat{\textbf{t}}^\prime,t)$ 
as:
\begin{equation}\label{expansion}
\psi(\textbf{r}^\prime,\hat{\textbf{t}}^\prime,t)=\psi(\textbf{r},\hat{\textbf{t}}^\prime,t)+(\textbf{r}^\prime-\textbf{r})\cdot\partial_{\textbf{r}}\psi(\textbf{r},\hat{{\bf t}}^\prime,t)+\cdots,
\end{equation}
the leading order terms will read as:
\begin{equation}\label{mean}
\overline{{\rm U}}^S(\textbf{r},\hat{{\bf t}},t) =-\frac{4}{3}\pi\ell_c^3 U_0 k_{\rm B}T\left(1+\frac{1}{10}\ell_c^2\,\nabla^2+\cdots\right)(\rho\,\hat{\textbf{t}}\cdot\textbf{P}).
\end{equation}
Now the short-range contributions will read as:
\begin{eqnarray}\label{TA}
&&\overline{\textbf{V}}^S(\textbf{r},\hat{{\bf t}},t)=\frac{4}{3}\pi \ell_c^3 U_0\,\textbf{D}\cdot\boldsymbol\nabla\left(\rho\,\hat{\textbf{t}}\cdot\textbf{P}\right)+\cdots,\nonumber\\
&&\overline{\boldsymbol\Omega}^S(\textbf{r},\hat{{\bf t}},t)=\frac{4}{3}\pi \ell_c^3 U_0 \,D_R\boldsymbol\nabla_R\left(\rho\,\hat{\textbf{t}}\cdot\textbf{P}\right)+\cdots.
\end{eqnarray}

Long-range contributions can also be obtained by inserting  (\ref{T}) and (\ref{R}) into equations (\ref{mft}) and (\ref{mfr}).  In terms of their components, the mean field long-range interactions  can be written as:
\begin{eqnarray}\label{meantr}
\overline{V}_i^L(\textbf{r},\hat{{\bf t}},t)=b_1T_i^1(\textbf{r},t)+b_2T_i^2(\textbf{r},t)+b_3T_{il}^3(\textbf{r},t)\,\hat{t}_l,
\end{eqnarray}
\begin{eqnarray}\label{meanrot}
\overline{\Omega}_i^L(\textbf{r},\hat{{\bf t}},t)=b_1T_{il}^4(\textbf{r},t)\,\hat{t}_l-b_2T_{il}^5(\textbf{r},t)\,\hat{t}_l+b_4T_{ilm}^6(\textbf{r},t)\,\hat{t}_l\hat{t}_m,
\end{eqnarray}
where summation over repeated indices is assumed and the coefficients are given by:
\begin{eqnarray}
b_1=\frac{261}{56} \,a\,\ell\,v_0\,\delta,\qquad b_2=-\frac{18}{35}\,\ell^3\,v_0\left({2+\delta}\right),\qquad b_3=\frac{5}{2}\,b_2,\nn\\
b_4=-\frac{45}{4}\ell^3\,v_0\left({2-\delta}\right).
\end{eqnarray}
Functions  $\textbf{T}^1,\cdots,\textbf{T}^6$ appeared  in (\ref{meantr}) and (\ref{meanrot})  
are functions of  position and their detailed structures are given by:
\begin{eqnarray}
T_i^1(\textbf{r},t)=\int \rmd\textbf{r}^\prime\,\frac{\hat{R}_i}{R^2}\,M_{jk}(\hat{\textbf{R}})\,\rho(\textbf{r}',t)\,N_{jk}(\textbf{r}',t),
\end{eqnarray}
\begin{eqnarray}
T_i^2(\textbf{r},t)=\int \rmd\textbf{r}^\prime\,\frac{1}{R^3}\,M_{ij}(\hat{\textbf{R}})\,\rho(\textbf{r}',t)\,P_j(\textbf{r}',t),
\end{eqnarray}
\begin{eqnarray}
T_{il}^3(\textbf{r},t)=\int \rmd\textbf{r}^\prime \, \partial_l \left(\frac{M_{jk}(\hat{\textbf{R}})}{R^3}\,R_i\right)\,\rho(\textbf{r}',t)\,N_{jk}(\textbf{r}',t),
\end{eqnarray}
\begin{eqnarray}
T_{il}^4(\textbf{r},t)=\int \rmd\textbf{r}^\prime\, \frac{\hat{R}_i}{R^3}\,M_{jkl}(\hat{\textbf{R}})\,\rho(\textbf{r}',t)\,N_{jk}(\textbf{r}',t),
\end{eqnarray}
\begin{eqnarray}
T_{il}^5(\textbf{r},t)=\int \rmd \textbf{r}^\prime\,\frac{1}{R^4}\,M_{ijl}(\hat{\textbf{R}})\,\rho(\textbf{r}',t)\,P_j(\textbf{r}',t),
\end{eqnarray}
\begin{eqnarray}
T_{ilm}^6(\textbf{r},t)=\int \rmd\textbf{r}^\prime \, \frac{\hat{R}_i}{R^4}\,M_{jklm}(\hat{\textbf{R}})\,\rho(\textbf{r}',t)\,N_{jk}(\textbf{r}',t).
\end{eqnarray}
In next sections, we will use the above results and study the dynamics of a suspension in the continuum 
limit.
\section{Continuum description}\label{sec4}
Now we can calculate the dynamical equations for the hydrodynamic continuum fields. Starting from 
equation (\ref{spd}), multiplying both sides by powers of ${\hat {\bf t}}$ and integrating over solid 
angle spanned by ${\hat {\bf t}}$, 
 we can obtain the equations that govern the dynamics of density, polarization and nematic order 
parameter.  Results of such calculations  can be written as:
\begin{eqnarray}\label{density0}
\partial_t\rho=-v_0 \boldsymbol\nabla . ( \rho \textbf{P} )+D_1 \nabla^2 \rho+D_2\,\partial_i\partial_j( \rho N_{ij})+\dot{\rho}^L+\dot{\rho}^S,
\end{eqnarray}
\begin{eqnarray}\label{polar0}
\partial_t(\rho P_i)=&-v_0\partial_j\left(\rho N_{ij}\right)-\frac{1}{3}v_0\partial_i \rho -2D_R\rho P_i\nn\\
 &+\partial_j\left(\frac{2}{5}D_2\,\partial_i(\rho P_j)+D_3\partial_j\left(\rho P_i\right)\right)+\dot{P}_i^L+\dot{P}_i^S,
\end{eqnarray}
\begin{eqnarray}\label{nematic0}
\partial_t (\rho N_{ij})=&-\frac{1}{5}v_0\left[\partial_i(\rho P_j)+\partial_j(\rho P_i)\right]+\frac{2}{15}v_0\,\delta_{ij}\boldsymbol\nabla\cdot(\rho\textbf{P})\nn\\
&-6D_R\rho N_{ij}+\frac{2}{15}D_2\left(\partial_i\partial_j-\frac{1}{3}\delta_{ij}\nabla^2\right)\rho\nn\\
&+\frac{2}{7}D_2 \partial_k\left(\partial_i(\rho N_{jk})+\partial_j(\rho N_{ik})-\frac{2}{3}\delta_{ij}\partial_l(\rho N_{kl})\right)\nn\\
&+D_4 \nabla^2(\rho N_{ij}) +\dot{N}_{ij}^L+\dot{N}_{ij}^S.
\end{eqnarray}
As one can see, in addition to the diffusion and swimmer's activity  terms, the terms proportional to $v_0$, there are contributions from interactions. 
Contributions from long-range and short-range interactions are collected in terms that are denoted by  superscripts 
$L$ and $S$ respectively ($\dot{\rho}^L$, $\dot{\rho}^S$   and {\it etc.}).  
To keep the continuity of text, we put these interaction terms in \ref{app2}.
Effective diffusion coefficients appeared in the above equations  are defined as:
\begin{eqnarray}
D_1=\frac{1}{3}\left(D_\parallel+2D_\perp\right),~~~
D_2=D_\parallel-D_\perp,\nn\\
D_3=\frac{1}{5}\left(D_\parallel+4D_\perp\right),~~~
D_4=\frac{1}{7}\left(D_\parallel+6D_\perp\right).
\end{eqnarray}

\subsection{Steady state solutions}\label{sec6}
\begin{figure}[htb]
\centering
\includegraphics[width=8cm]{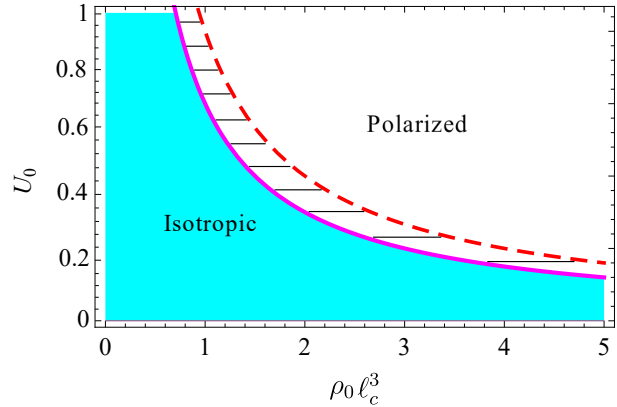}
\caption{
Phase diagram showing  possible thermodynamic phases for a suspension of swimmers. Isotropic and polarized phases are 
separated by a solid line in a space given by $U_0$ (strength of short-range interaction) and $\rho_0 \ell_c^3$ (density of swimmers). Taking into account long-wavelength fluctuations, calculations show  that both phases are unstable.  
Part of polarized phase denoted by dashed lines, shows the states that are stable with respect to splay fluctuations. These states are unstable with respect to bend fluctuations.}
\label{fig3}
\end{figure}

Here we seek for steady state uniform solutions of the above  dynamical equations 
for continuum fields. Terms corresponding to  long-range interactions and swimmer's activity,  do not contribute in the uniform steady state 
solutions.  Steady states are solutions to the following equations:  
\begin{eqnarray}
\partial_t\rho=0,
\end{eqnarray}
\begin{eqnarray}
\partial_t(\rho P_i)=-2D_R\rho P_i+\frac{4}{3}\pi D_R\ell_c^3 U_0\rho^2 \left(\frac{2}{3}P_i - P_jN_{ij}\right),
\end{eqnarray}
\begin{eqnarray}
\partial_t (\rho N_{ij})=-6D_R\rho N_{ij}+\frac{8}{5}\pi D_R\ell_c^3 U_0\rho^2 \left(P_i P_j-\frac{P^2}{3}\delta_{ij}\right).
\end{eqnarray}
As a result of the above equations,  we realize that there are two different homogenous steady state phases in our system. The first phase, denoted by phase $I$, is an isotropic phase and  defined by:
\begin{eqnarray}
\rho^{I}=\rho_0,\,~~~~~ \textbf{P}^{I}=\textbf{0},\,~~~~~ \textbf{N}^{I}=\textbf{0}.
\end{eqnarray}
In this phase, all swimmers are distributed uniformly in the fluid and move randomly without any 
preferred direction. 
Increasing the density, we see that beyond a critical density  $\rho_0>\rho_c=9/(4\pi\ell_c^3U_0)$, a homogeneous polarized state appears. This phase is denoted by phase $P$ and defined by: 
\begin{eqnarray}
\rho^{P}=\rho_0,\,~~~~~\textbf{P}^{P}=\textbf{P}^\infty,\,~~~~~\textbf{N}^{P}=\textbf{N}^\infty. 
\end{eqnarray}
In this polarized phase, swimmers are distributed uniformly and move in a preferred direction.
Steady state polarization and nematic order parameter in the polar phase are given by:
\begin{eqnarray}
\textbf{P}^{\infty}=\sqrt{\frac{15}{4\pi\ell_c^3\rho_0 U_0}\left(1-\frac{9}{4\pi\ell_c^3\rho_0 U_0}\right)}\,\hat{\textbf{n}},
\end{eqnarray}
\begin{eqnarray}
\textbf{N}^{\infty}=\left(1-\frac{9}{4\pi\ell_c^3\rho_0 U_0}\right)\left(\hat{\textbf{n}}\hat{\textbf{n}}-\frac{{\textbf{I}}}{3}\right),
\end{eqnarray}
where  $\hat{\textbf{n}}$ denotes the direction of broken symmetry. 
Figure~\ref{fig3}, shows a phase diagram in a space characterized by $U_0$ and $\rho_0\ell_{c}^{3}$. 
Appearance  of the ordered phase is a direct consequence of the  short-range (alignment)  interaction 
between the swimmers. As it is apparent from the  equations, long-range interactions alone, are not able to induce any ordered state in bulk \cite{Marchetti2}. It is shown very recently that short-range 
hydrodynamic interactions in symmetric squirmers are also able to induce a polar state \cite{Liverpoolsquiremer}. 
\subsection{Stability of Isotropic state}\label{sec7}
\label{stabilityanalysisI}
In addition to  existence of  steady state phases, their stability is important to analyze.  Thermal or non-thermal fluctuations can destabilize the above steady state phases. In this section, we  study the  stability of the steady state solutions.

To study the stability of  isotropic phase, we   add small fluctuations to the corresponding fields 
of the isotropic state and investigate their dynamics: 
\begin{eqnarray}\label{iso-rho}
\rho(\textbf{r},t)=\rho_0+\delta\rho(\textbf{r},t),
\end{eqnarray}
\begin{eqnarray}
\textbf{P}(\textbf{r},t)=\textbf{0}+\delta\textbf{P}(\textbf{r},t),
\end{eqnarray}
\begin{eqnarray}\label{iso-N}
\textbf{N}(\textbf{r},t)=\textbf{0}+\delta\textbf{N}(\textbf{r},t).
\end{eqnarray}
Using the dynamical equations obtained in the above section, we can obtain the 
evolution equations of these fluctuating fields. To linearize the equations, we  introduce spatial Fourier transformation 
as: 
\begin{eqnarray}
\check{f}(\textbf{k})=\int\rmd\textbf{r}\,\rme^{\rmi\textbf{k}\cdot\textbf{r}}\,f(\textbf{r}).
\end{eqnarray}
In \ref{app2}, we have shown how a typical term in the dynamical equation can be linearized. 
Repeating the same procedure for all other terms, we can arrive at the following equations that describe the 
linearized dynamics of the fluctuations around the isotropic phase:
\begin{eqnarray}\label{density}
\partial_t \delta \check{\rho}={\rm i}v_0\rho_0 k_i\delta \check{P}_i-\frac{8\pi{\rm i}}{9}b_2\rho_0^2 k_i \delta \check{P}_i-D_1k^2\delta \check{\rho}-\rho_0 D_2k_ik_j\delta\check{N}_{ij},
\end{eqnarray}
\begin{eqnarray}\label{polarization}
\partial_t \delta\check{P}_i=&{\rm i}v_0k_j\delta\check{N}_{ij}+{\rm i}\frac{v_0}{3\rho_0}\delta\check{\rho}k_i-\frac{8\pi{\rm i}}{45}b_3\rho_0 k_j\delta\check{N}_{ij}-\frac{2}{5}D_2k_j\delta\check{P}_jk_i-D_3k^2\delta\check{P}_i\nonumber\\
&+\frac{4}{9}\pi\ell_c^3 U_0 \rho_0 \left(\frac{1}{5}D_2 [2k_j\delta\check{P}_jk_i+k^2\delta \check{P}_i]+D_\perp k^2\delta \check{P}_i\right)-2D_R\delta\check{P}_i\nonumber\\
&+\frac{8}{9}\pi D_R\ell_c^3 U_0\rho_0 \delta\check{P}_i-\frac{32\pi{\rm i}}{225}b_4\rho_0 k\left(-\hat{k}_j\hat{k}_k\delta\check{N}_{jk}\hat{k}_i+\frac{14}{35}\delta\check{N}_{ik}\hat{k}_k\right)
\end{eqnarray}
and
\begin{eqnarray}\label{nematic}
\partial_t \delta\check{N}_{ij}=&\frac{2}{5}{\rm i}v_0\left(\frac{1}{2}[k_i\delta\check{P}_j+k_j\delta\check{P}_i]-\frac{1}{3}k_k\delta\check{P}_k\delta_{ij}\right)-\frac{2}{15\rho_0}D_2k_ik_j\delta\check{\rho}\nonumber\\
&+\frac{2}{45}D_2k^2\delta_{ij}\frac{\delta\check{\rho}}{\rho_0}-6D_R\delta\check{N}_{ij}-D_4 k^2\delta\check{N}_{ij}-\frac{2}{7}D_2\Bigg(k_zk_i\delta\check{N}_{jz}\nonumber\\
&+k_zk_j\delta\check{N}_{iz}-\frac{2}{3}\delta_{ij}k_kk_l\delta \check{N}_{kl}\Bigg)+\frac{\rho_0}{5}\Bigg(\frac{8\pi}{3}b_1[2\hat{k}_k\hat{k}_l\delta\check{N}_{kl}\hat{k}_i\hat{k}_j\nonumber\\
&-\delta\check{N}_{ik}\hat{k}_k\hat{k}_j-\delta\check{N}_{jk}\hat{k}_k\hat{k}_i+\frac{2}{5}\delta\check{N}_{ij}]+\frac{8\pi{\rm i}}{15}b_2[5k_k\delta\check{P}_k\hat{k}_i\hat{k}_j\nonumber\\
&-\delta\check{P}_ik_j-\delta\check{P}_jk_i-k_k\delta\check{P}_k\delta_{ij}]\Bigg).
\end{eqnarray}
These  coupled equations, govern the dynamics of fluctuations. 
As the analysis of above coupled equations is not simple, we can use different approximations to understand  
 physical mechanisms of possible instabilities. 

As a first approximation and at times longer than time scale of rotational diffusion ($t\gg D_R^{-1}$), we can neglect 
the dynamics of $\delta\check{P}_{i}$ and $\delta\check{N}_{ij}$ in equations (\ref{polarization}) and (\ref{nematic}). 
Solving the simplified equations for polarization and  nematic fluctuations  ($\partial_t\delta\check{P}_{i}=\partial_t\delta\check{N}_{ij}\rightarrow 0$),  we can substitute them in equation (\ref{density}) and keep leading order powers of wave vector $k$. This will result  an effective diffusion equation for density 
fluctuations as:
\begin{eqnarray}
\partial_t \delta \check{\rho}=-D_{eff}k^2\delta\check{\rho},
\end{eqnarray}
where the effective diffusion constant is given by:
\begin{eqnarray}\label{dif}
D_{eff}=D_1+\frac{v_0^2}{D_R(3-\frac{4}{3}\pi\ell_c^3\rho_0U_0)}\Bigg[\frac{1}{2}+\frac{8\pi\rho_0\ell^3}{35}\Bigg].
\end{eqnarray}
Since in the isotropic phase $\ell_c^3\rho_0U_0<9/4\pi$, the above effective diffusion coefficient is always positive and it is greater than diffusion coefficient of a  brownian self-propelled rod, $D_1+v_0^2/6D_R$ \cite{Howse}.  As a result of 
positivity of $D_{eff}$,  density fluctuations  damp and the isotropic state is   
stable if we neglect polarization and nematic fluctuations. 

It is a well known fact that for an active brownian particle, orientational fluctuations increase the 
translational diffusion by a term proportional to $v_0^2/D_R$, but what is new here is the effect of hydrodynamic interactions. In the above result,    
the second term in bracket, $({8\pi}/{35}){\rho_0\ell^3}$, which is due to hydrodynamic interaction, shows that  the hydrodynamic 
interaction  speeds up the diffusion process.  The increase in diffusion due to the hydrodynamic 
interaction is proportional to the density of swimmers and also the size of an individual swimmer.

\begin{figure}
\centering
\includegraphics[width=12cm]{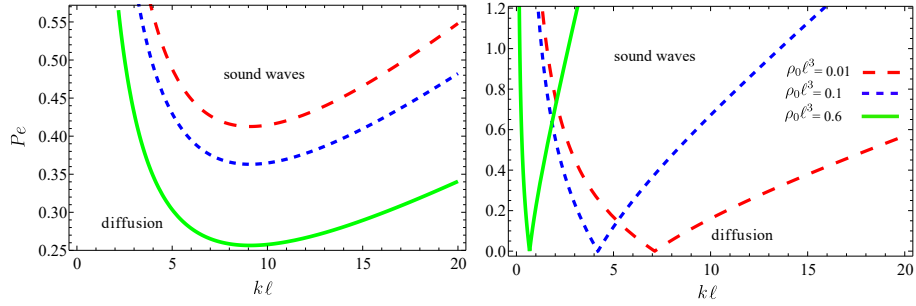}
\caption{  For a system with given 
$\rho_0\ell^3$ and depending on the value of  P{\'e}clet number
$Pe=\frac{v_0\,\ell}{D_\parallel}$,  a fluctuating mode with wave vector $k\ell$, can propagate with diffusion 
or sound wave mechanism. Lines show the boundary between these two different bahavior. Left graph is for a system that has only long-range interactions but the 
right graph shows the results for a system that has both short- and long-range interactions.
 Numerical values we have used are: $\delta=0.1$, $a/\ell=0.1$, $\ell_c/\ell=1$ and $U_0=1$.}
\label{fig5}
\end{figure}

To have more insights on the fluctuations in the isotropic phase,  we  study the dispersion relation 
for  hydrodynamic modes in the system.  In this case, we do not base our approximation on  
neglecting  the dynamics of polarization and 
nematic order parameter from above coupled equations. It is apparent from equations (\ref{polarization}) and (\ref{nematic}) that the nematic fluctuations are coupled to density fluctuations in higher powers of 
wave vector.  We are interested in the long-wavelength fluctuations so, as an another approximation we may discard nematic fluctuations 
($\delta\check{N}_{ij}\rightarrow 0$) and  consider only the coupled dynamics of 
density and polarization fluctuations. Assuming a time dependent form for the fluctuations as:
\begin{eqnarray}
\delta\check{\rho}, \delta \check{P}_i\sim e^{\chi(k) \,t},
\end{eqnarray}
we can study their coupled dynamics and obtain a   
dispersion relation like $\chi=\chi(k)$. Up to the leading orders of $k$, 
dispersion relations read:
\begin{eqnarray}
\chi_\pm=\frac{1}{54}\Bigg[&-9\left(3-\frac{4}{3}\pi\ell_c^3\rho_0U_0\right)(2D_R+D_5k^2)-27D_1k^2\nonumber\\
&\pm\Bigg[\left(9\left(3-\frac{4}{3}\pi\ell_c^3\rho_0U_0\right)(2D_R+D_5k^2)+27D_1k^2\right)^2\nonumber\\
&\qquad +108\Bigg(-9D_1k^2\left(3-\frac{4}{3}\pi\ell_c^3\rho_0U_0\right)(2D_R+D_5k^2)\nonumber\\
&\qquad -9v_0^2k^2\left(1+\frac{16\pi}{35}\rho_0 \ell^3(2+\delta)\right)\Bigg)\Bigg]^{1/2}\Bigg],
\end{eqnarray}
where $D_5=1/5\,(3D_\parallel+2D_\perp)$. 
As we expected from previous discussion, for $\ell_c^3\rho_0U_0<9/4\pi$, both $\chi_+$ and $\chi_-$ are negative, reflecting the fact that the isotropic phase is always stable.   It should be also noted that albeit the two modes have always negative real values, but if the self-propulsion speed of the swimmers is greater than a threshold value, they will have an imaginary part. As a result of this imaginary part, fluctuations of density and polarization will decay with a propagating mechanism and propagating sound waves 
 will  appear in the system \cite{Marchetti}.  In terms of P{\'e}clet number $Pe=({v_0\,\ell})/{D_\parallel}$ and 
 dimensionless wave vector $k\ell$, figure \ref{fig5}, shows  the regions where these waves can propagate. For an 
 intermediate $k\ell$, density waves appear at larger $Pe$. As seen in 
 figure \ref{fig5} (left), taking into account only long-range part of the interactions,  increasing the density 
 will decrease the threshold $Pe$ above which propagating waves appear. Taking into account both 
 long- and short-range interactions in figure \ref{fig5} (right), we see that  smaller densities of swimmers 
 have a wider region for density waves. 
Interestingly, all these results are valid for both pushers and pullers.

Above approximations show that density and polarization fluctuations are not able to induce any instability in the 
isotropic phase. To see how the nematic fluctuations can provide mechanism for instability, we can 
study its dynamics separately. 
 Arranging  the right hand side of equation (\ref{nematic}) in powers of $k$, we can study the nematic 
 fluctuations at long-wavelength limit. Keeping the  leading order term, we find that the nematic fluctuations are decoupled from
 density and polarization as:
 \begin{eqnarray}
\partial_t \delta N_s=\left(-6D_R+\frac{16 \pi}{75}\rho_0 b_1\right)\delta N_s,
\end{eqnarray}
\begin{eqnarray}
\partial_t \delta \textbf{N}_b=\left(-6D_R-\frac{8 \pi}{25}\rho_0 b_1\right)\delta \textbf{N}_b,
\end{eqnarray}
where nematic fluctuations are decomposed into their  splay, $\delta N_s=\hat{\textbf{k}}\cdot\delta\check{\textbf{N}}\cdot\hat{\textbf{k}}$, and bend $\delta \textbf{N}_b=\hat{\textbf{k}}\cdot\delta\check{\textbf{N}}\cdot\left(\textbf{I}-\hat{\textbf{k}}\hat{\textbf{k}}\right)$ components.  
Coefficient $b_1$ is proportional to the asymmetry parameter $\delta$ and for pullers (pushers) it is positive (negative). 
This shows that, if the density of swimmers is greater than a value given by $\rho_{ins}\propto D_R/|b_1|$, 
splay (bend) perturbations in nematic tensor can destabilize an isotropic suspension of pullers (pushers). Such resolution 
in the instability of pushers and pullers will be seen at the next section.

\subsection{Stability of Polar state}
\label{stabilityanalysisII}
To  study the  stability of polar phase, we assume that the density of swimmers is larger than $\rho_c$, 
so that the polarized phase has been established. Then we study the dynamics of fluctuations around 
the polarized state. Denoting by $\hat{\mathbf{n}}$, the direction of polarization, we suppose 
that the order parameter has a constant value but its direction  fluctuates. In this case, hydrodynamic fields can be written as:
\begin{eqnarray}
\rho=\rho_0+\delta\rho,
\end{eqnarray}
\begin{eqnarray}
\mathbf{P}=P^\infty\, \left(\hat{\mathbf{n}}_0+\delta\mathbf{n}\right),
\end{eqnarray}
\begin{eqnarray}
\mathbf{N}=N^\infty\left((\hat{\mathbf{n}}_0+\delta\mathbf{n})(\hat{\mathbf{n}}_0+\delta\mathbf{n})-\frac{\mathbf{I}}{3}\right).
\end{eqnarray}
Where we have assumed $\hat{\textbf{n}}=\hat{\textbf{n}}_0+\delta\textbf{n}$, with $\hat{\textbf{n}}_0$ is the 
average direction of polarization in  system. 
Furthermore  for small fluctuations we have 
$\hat{\mathbf{n}}_0\cdot\delta {{\bf n}}=0$. 
Using the above definitions, we can linearize equations (\ref{density0}) and (\ref{polar0}) and obtain evolution equations for density and director fluctuations. Since we are doing our calculations in the Fourier space, the angle between wave vector $\mathbf{k}$ and director $\hat{\mathbf{n}}_0$ will emerge in the linearized equations. 
To simplify the analysis, we decompose the fluctuations into bend and splay distortions.  
Splay distortion is a fluctuation with  $\nabla\cdot {\bf P}\neq 0$ ($\nabla\cdot \delta {\bf n}\neq 0$) and for  bend fluctuations 
${\bf P}\times(\nabla\times {\bf P})\neq 0$ ($\nabla\times \delta {\bf n}\neq 0$). 
Decomposing  the wave vector ${\bf k}$ into its parallel 
and perpendicular  components as: 
${\bf k}=({\bf k}\cdot\hat{{\bf n}}_0)\hat{\mathbf{n}}_0+{\bf k}_\perp$, we can see that 
for bend (splay) fluctuations only the parallel (perpendicular) component of the wave vector contributes.
These  two modes of fluctuations are independent and this allows us to  study them separately. 

To study the bend  fluctuations, we can  set  $\mathbf{k}=k\hat{\mathbf{n}}_0$ and study the dynamics
of fluctuations. Using a linearization procedure similar to what we have used at previous section and 
keeping terms up to second order of wave vector, we can arrive at  the following equations for 
bend fluctuations:
\begin{eqnarray}
\partial_t\delta\check{\rho}_{\rm b}=\left(\rmi\,\chi_\rho^{Im}+\chi_\rho^{Re}\right)\delta\check{\rho}_{\rm b},
\end{eqnarray}
\begin{eqnarray}
\partial_t\delta\check{\mathbf{n}}_{\rm b}=\left(\rmi\,\chi_n^{Im}+\chi_n^{Re}\right)\delta\check{\mathbf{n}}_{\rm b},
\end{eqnarray}
where the imaginary and real parts are given by:
\begin{eqnarray}\label{rerho}
\chi_\rho^{Re}=-\frac{16\pi}{9}\rho_0\,N^\infty\,b_1+k^2\left(-D_1+N^\infty\,D_6\right),
\end{eqnarray}
\begin{eqnarray}\label{imrho}
\chi_\rho^{Im}=P^\infty\,k\left(v_0-\frac{8\pi}{9}\rho_0\,b_2+\frac{64\pi}{45}\rho_0\, N^\infty\,b_3\right),
\end{eqnarray}
\begin{eqnarray}\label{ren}
\chi_n^{Re}=-\frac{136\pi}{75}b_1\rho_0N^\infty+\frac{4}{35}k^2\,D_2\left(\frac{4}{3}\pi\ell_c^3\rho_0U_0-3\right),
\end{eqnarray}
\begin{eqnarray}\label{imn}
\chi_n^{Im}=\frac{k}{\rho_0P^\infty}\Bigg(&v_0\rho_0N^\infty-\frac{8\pi}{45}\rho_0^2b_3N^\infty(1-N^\infty)\nonumber\\
&-\frac{4\pi}{25}b_2\rho_0^2P^{\infty 2}-b_4\rho_0^2N^\infty\left[\frac{2368}{(105)^2}N^\infty+\frac{192}{875}\right]\Bigg),
\end{eqnarray}
with $D_6=1/3\,(7D_\parallel+8D_\perp)$. As it is seen from the above equations, fluctuations of  density and 
polarization  are decoupled for the case of bend distortions. Both of modes show that sound-like density waves can propagate in the system; regions with dense ordered population of particles propagating in a disordered background. Propagation of these waves is a signature of Vicsek-type flocking models \cite{Peruani3,Schaller}. 

To analyze the stability of polar state against bend fluctuations, let us consider two cases,
first: without hydrodynamic interactions and second: with hydrodynamic interactions. 
The terms proportional to $b_i$ in the above equations, originate from long-range hydrodynamic interactions. 
In the absence of hydrodynamic interactions where $b_i=0$, real parts in both of the above 
equations are of order $k^2$, revealing the diffusing nature of the fluctuations. 
Moreover, under these conditions density fluctuations are damped for $\ell_c^3\rho_0U_0<\frac{3}{4\pi}\frac{7D_\parallel+8D_\perp}{2(D_\parallel+D_\perp)}\sim 0.9$ (we used numerical values as: $\delta=0.1$ and $a/\ell=0.1$). These states are denoted by   dashed region in  figure \ref{fig3}.  Beyond this region and for 
$\ell_c^3\rho_0U_0>0.9$, density fluctuations can grow and form  clusters of swimmers.
Considering the polarization fluctuations, we can see that for  $\ell_c^3\rho_0U_0>9/4\pi$, such 
fluctuations can always grow and make the polar state unstable.

Taking into account both short- and long-range interactions and in the limit of long-wavelength fluctuations 
($k\rightarrow 0$), the terms that are proportional to $b_1$ in (\ref{rerho}) and (\ref{ren}), 
are the most important terms that determine the instability criterion.
Recalling the fact that  $b_1\propto \delta$, we see that for  pullers ($\delta>0$) density and director fluctuations diminish, but they diverge  for pushers ($\delta<0$).    The growth of bend fluctuations 
destabilizes any polar order in a  suspension of pushers \cite{Ramaswamy4}.

To study the role of splay fluctuations, we set  
${\bf k}=\mathbf{k}_\perp=k\hat{\mathbf{n}}_\perp$, 
with $\hat{\mathbf{n}}_\perp\cdot\hat{\mathbf{n}}_0=0$. 
For splay distortions,  fluctuations of density and director   are always coupled to each other and they   
obey the following equations:
\begin{eqnarray}\label{drho}
\partial_t\delta\check{\rho}_s=H_{11}\delta\check{\rho}_s+H_{12}\,\delta\check{n}_{\rm s},
\end{eqnarray}
\begin{eqnarray}\label{dn}
\partial_t\delta\check{n}_{\rm s}=H_{21}\delta\check{\rho}_s+H_{22}\,\delta\check{n}_{\rm s},
\end{eqnarray}
where
\begin{eqnarray}
H_{11}= \frac{8}{9} \pi b_1 \rho_0 N^\infty+k^2 \left(-D_1+N^\infty D_7\right),
\end{eqnarray}
\begin{eqnarray}
 H_{12}=\rmi k\rho_0 P^\infty \left(v_0 \rmi - \frac{8}{9} \pi b_2 \rho_0 - \frac{2}{5} \pi b_3 \rho_0 N^\infty\right),
\end{eqnarray}
\begin{eqnarray}
 H_{21}=\frac{\rmi k}{\rho_0 P^\infty} \Bigg(&\frac{1}{3} v_0 (1-N^\infty)+\frac{32\pi}{135}\rho_0 b_3 N^\infty(\frac{2}{7}N^\infty-1)\nonumber\\
&-\frac{4\pi}{25}\rho_0 b_2 P^{\infty 2}-\frac{16\pi}{175}\rho_0 b_4 N^\infty(\frac{6}{5}+\frac{29}{63}N^\infty)\Bigg),
\end{eqnarray}
\begin{eqnarray}
H_{22}=\frac{64\pi}{75}b_1 \rho_0 N^\infty -\frac{3}{35}D_2k^2\left(\frac{4}{3}\pi\ell_c^3\rho_0U_0-3\right),
\end{eqnarray}
with $D_7=1/3\,(4D_\parallel+11D_\perp)$ and $\delta\check{n}_{\rm s}=\hat{\mathbf{n}}_\perp\cdot\delta\check{\mathbf{n}}$. 
By calculating  eigenvalues of  matrix $\textbf{H}$,  we will obtain two dispersion relations for the fluctuation 
spectrum.  In the absence of hydrodynamic interactions, the spectrum of fluctuation  has a simpler form: 
\begin{eqnarray}
\label{nonhyd}
\chi_\pm=\pm \frac{\rmi kv_0}{\sqrt{\frac{4}{3}\pi\ell_c^3\rho_0 U_0}}+k^2\left(D_8-\frac{3D_{9}}{4\pi\ell_c^3\rho_0 U_0}-\frac{2}{35}\pi\ell_c^3\rho_0 U_0D_2\right),
\end{eqnarray}
where $D_8=2/35(11D_\parallel+24D_\perp)$, $D_{9}=1/2(4D_\parallel+11D_\perp)$. 
Real part of this relation is negative for $\ell_c^3\rho_0 U_0<0.9$, reflecting the fact that  in 
the absence of hydrodynamic interactions, splay fluctuations  will decay to zero when $9/4\pi<\ell_c^3\rho_0U_0<0.9$, (dashed region in figure \ref{fig3}). But for  $\ell_c^3\rho_0U_0>0.9$, above the red dashed line in figure \ref{fig3}, splay fluctuations diverge, hence make the 
polarized state unstable. 

If we  consider the contributions from hydrodynamic interactions and in long-wavelength limit, the 
dispersion relations for the splay fluctuations read as:  
\begin{eqnarray}
\chi_+=\frac{4}{5}\pi \rho_0 N^\infty b_1+{\cal O}(k)^2,~~~\chi_{-}=\frac{1}{5}\pi \rho_0 N^\infty b_1+{\cal O}(k)^2.
\end{eqnarray}
As for the bend fluctuations, the sign of  $b_1\propto \delta$ determines the 
criterion for instability. For a suspension of pullers ($b_1>0$) fluctuations grow but  for pushers ($b_1<0$) fluctuations damp to zero. So an ordered suspension of pullers becomes unstable by the growth of splay fluctuations. 


\begin{figure}[htbp]
\centering
\includegraphics[width=10cm]{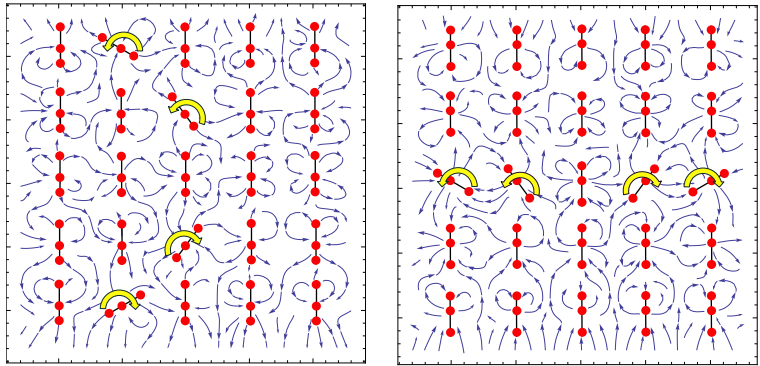}
\caption{
Flow patterns for  demonstrating bend (left) and splay (right) distortions in polarized phase. An ordered suspension of pushers (pullers) is unstable due to the growth of bend (splay) fluctuations. Small arrows show the direction of flow field in the inter particle regions and large arrows show the flow field calculated at the position of distorted swimmers.
}
\label{bendsplay}
\end{figure}

\section{Summary and Discussion}
\label{sec5}
In this article, we have started from a microscopic model for a hydrodynamic micro-swimmer and 
have derived its average dynamical characteristics such as  velocity and  force distribution. The swimmer that 
we have started with, is able to model both pushers and pullers. We have  shown that a set of analytical 
expressions can be obtained for the long-range interactions between two swimmers.
Extending the system to a three dimensional dilute suspension of  swimmers and considering   two body interaction 
 between swimmers, we have developed a  continuum description that can capture  thermodynamic 
 properties of the suspension. Furthermore, we assumed that in addition to long-range interactions, there 
 is a short-range interaction that can align the nearby swimmers.  

What we aimed in this article was to investigate the role of interactions in long-wavelength 
instabilities of the suspension. Isotropic phase and a symmetry broken  polar phase, 
are two possible thermodynamic phases of the system.  Depending on the density of swimmers, 
at low density of swimmers, the system is in isotropic phase and increasing the density 
will lead the system to a polar phase. 
In a system with hydrodynamic interactions, both of the above  phases are unstable with respect to 
long-wavelength fluctuations. 
It is the long-range interaction that initiates the instability in an interacting suspension. 
Our results are compatible 
with the well known results of phenomenological models that state the origin of instability. 
Decomposing the nematic distortions  into  bend and splay fluctuations, we show that for a suspension of 
pushers, bend fluctuations mediate the instability and for a suspension of pullers it is the splay fluctuation 
that initiates the instability. 
Intuitional arguments can help to have more insights on the instability of polar phase.
Figure \ref{bendsplay}(left), shows a regular collection of pushers with polar order.   A small bend fluctuation is introduced to this collection by distorting the director of five selected 
swimmers. For a  regular system, fluid velocity due to the other swimmers averages to zero at 
the position of each swimmer, but for the distorted case shown in this figure, fluid 
velocity has nonzero value at the position of 
distorted swimmers. As shown by large arrows, the velocity streamlines at the position of distorted 
swimmers are in the direction that tend to increase the initial distortions and destroy the initial regular state.   
Figure \ref{bendsplay}(right),
shows the case for pullers with a small splay fluctuation. For pullers, by applying a small splay 
fluctuation, the system will tend to increase it and destroy the polar order. 

Another interesting feature in active nematic is the appearance of bands in polar state. In   
symmetry broken polar phase, density waves will appear. Imaginary parts appeared in equations (\ref{imrho}), 
(\ref{imn}) and (\ref{nonhyd}),  reflect this fact. Interestingly, in the case of splay fluctuations,  
a single group velocity for these traveling waves  is seen.
Finally, we should mention that all instabilities arized from hydrodynamic interactions, are for dipolar swimmers. For a 
collection of neutral swimmers with quadrupolar force distributions, the 
terms proportional to $b_1$ in equations (\ref{meantr}) and (\ref{meanrot}) do not contribute and all ordered phases are stable with respect to long-wavelength fluctuations. 

\section*{Acknowledgement}
Useful discussions with  M. C. Marchetti and K. Kruse are acknowledged.

\appendix

\section{Hydrodynamic center and diffusion coefficients for a rigid swimmer}
\label{app1}
Here we want to show how the hydrodynamic center and diffusion coefficients for a swimmer can 
be calculated. Let us consider a rigid swimmer composed of three spheres with equal radii $a$, linked linearly by two negligible diameter linkers. Labeling spheres by $f$, $m$ and $b$, the front linkage has 
a length given by $L^f=\ell$ and the back linkage has a length given by $L^b=\ell(1+\delta)$. 
Hydrodynamic center for this rigid system is a point around which the translational motion  
is independent from the rotational motion. 
As a result of symmetry, for our linear three linked spheres, the hydrodynamic center lies somewhere on the longer 
 linkage 
with a distance $x$ from the middle sphere. Hydrodynamic center is a geometrical concept and it is independent 
from  dynamics, but we can benefit any dynamical problem  to calculate it. 
Let us consider a dynamical problem that as a result of an external force, 
the hydrodynamic center moves linearly without any net rotation.   
With respect to hydrodynamic center, total torque should vanish: $xf_\perp^{m}+(\ell+x)f_\perp^f-\left(\ell(1+\delta)-x\right)f_\perp^b=0$ where, $\perp$ denotes the components of vectors 
perpendicular to the linkages. In addition to this condition, 
there is a set of linear equations that relates the forces and 
velocities as: ${\bf v}^{\alpha}=\sum_{\beta=f,m,b}{O}^{\alpha\beta}{\bf f}^\beta$ where ${O}$ denotes the 
Oseen's tensor. We can use this set of equations and find relations between perpendicular components of forces 
and velocities. Rigidity condition is  another equation that we must consider: 
$v^{f}_{\perp}=v^{m}_{\perp}=v^{b}_{\perp}$. Using the rigidity and force-velocity equations we can obtain 
relations for $f_\perp^{f}/f_\perp^{m}$ and $f_\perp^{b}/f_\perp^{m}$ and plugging them into the torque 
free condition, we can obtain the following result for $x$:  
\begin{eqnarray}
x=\frac{1}{3}\delta\left(\ell+\frac{7}{8}a\right).
\end{eqnarray}

\begin{figure}[htbp]
\centering
\includegraphics[width=6cm]{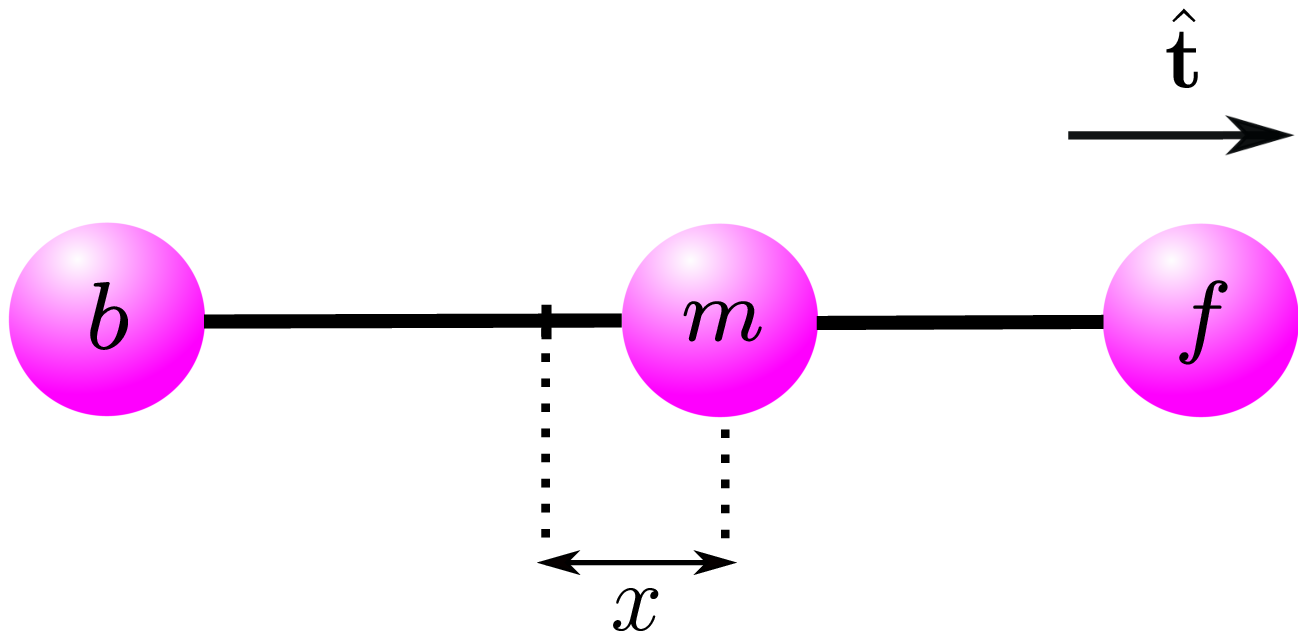}
\caption{
In an asymmetric three-sphere swimmer, hydrodynamic center is located at a distance $x$ from the middle sphere.}
\label{fig6}
\end{figure}

Having in hand the position of hydrodynamic center, we can calculate the translational and rotational 
diffusion coefficients. 
To obtain the translational diffusion coefficients, let us apply an external force ${\bf f}_T$ to the 
system and calculate the  linear velocity ${\bf v}_T$ that the  system will acquire. Then the translational diffusion
matrix ${\bf D}$ is defined by: $\textbf{v}_T=(k_{\rm B}T)^{-1}\textbf{D}\cdot\textbf{f}_T$. 
To calculate ${\bf D}$, one should note that in addition to force-velocity relations,  constraints of  total force ${\bf f}^f+{\bf f}^m+{\bf f}^b={\bf f}_T$ and 
rigidity 
${\bf v}^f={\bf v}^m={\bf v}^b={\bf v}_T$ should be considered. Solving these equations, we will have:
\begin{eqnarray}
(k_{\rm B}T)^{-1}D_{ij}=K(\ell)\,\hat{t}_i\hat{t}_j+K(2\ell)\,\left(\delta_{ij}-\hat{t}_i\hat{t}_j\right),
\end{eqnarray}
where
\begin{eqnarray}
K(\ell)=\frac{1}{18\pi\eta a}\Bigg[1+\frac{5}{2}\left(1-\frac{\delta}{2}\right)\left(\frac{a}{\ell}\right)+\mathcal{O}\left(\frac{a}{\ell}\right)^2\Bigg],
\end{eqnarray}
In terms of its parallel and perpendicular components, the diffusion coefficients are given by: 
${D}_\parallel=k_{\rm B}TK(\ell)$ and ${D}_\perp=k_{\rm B}TK(2\ell)$.

To calculate the rotational diffusion coefficient, we apply an external torque $\tau$ around 
the hydrodynamic center, then the system will rotate with angular velocity $\Omega$ around  that center with no translation for hydrodynamic center. Rotational diffusion can be calculated as: 
$\Omega=(k_{\rm B}T)^{-1}D_R\, \tau$. 
In this case, in addition to the linear force-velocity relations given by Oseen's tensor, we must 
consider the torque equation as 
$\tau=xf_\perp^{m}+(\ell+x)f_\perp^f-\left(\ell(1+\delta)-x\right)f_\perp^b$ and rigidity constraints as:
$v_\perp^{m}=x \Omega$, $v_\perp^f=(\ell+x)\Omega$
and $v_\perp^b=-\left(\ell(1+\delta)-x\right)\Omega$.
Solving  all these equations simultaneously, the final result reads as:
\begin{eqnarray}
D_R=\frac{k_{\rm B}T}{12\pi\eta a\,\ell^2}\Bigg[(1-\delta)-\frac{3}{8}\left(1-\frac{3}{2}\delta\right)\left(\frac{a}{\ell}\right)+\mathcal{O}\left(\frac{a}{\ell}\right)^2\Bigg].
\end{eqnarray}

\section{Details of  interaction terms}
\label{app2}
In this appendix we give the details of interaction contributions introduced in equations 
(\ref{density0}), (\ref{polar0}) and (\ref{nematic0}). Interaction contributions to the dynamics of density, polarization and nematic order parameter 
read as:
\begin{eqnarray}\label{hyd-rho}
\dot{\rho}^L=-\partial_i\left(\left[ b_1\,T_i^1+b_2\,T_{i}^2 \right]\,\rho+b_3\,T_{il}^{3}\,\rho\,P_l \right).
\end{eqnarray}

\begin{eqnarray}
\dot{\rho}^S=-\frac{4}{3}\pi\ell_c^3 U_0\partial_{i}\Bigg(\rho\,\partial_{j}(\rho P_k)\Bigg(&\frac{1}{5} D_2\left(\delta_{kj}P_i+\delta_{ik}P_j+\delta_{ij}P_k\right)\nn\\
&+D_\perp\delta_{ij}P_k\Bigg)\Bigg).
\end{eqnarray}

\begin{eqnarray}
\dot{P}_i^L= &- \partial_j \left((b_1\,T^{1}_j+b_2\,T^{2}_j) \, \rho\,P_i+ b_3\,T^3_{jl}\,\rho\,\left(N_{il}+\frac{\delta_{il}}{3}\right) \right)\nn\\
&+\frac{4}{5}\left(b_1T_{il}^4-b_2T_{il}^5\right)\rho P_l-\frac{1}{5}b_1T_{ll}^4\,\rho P_i+\frac{6}{7}b_4T_{ilm}^6\,\rho N_{lm}\nn\\
&-\frac{1}{5}\left(b_1T_{li}^4-b_2T_{li}^5\right)\rho P_l-\frac{2}{7}b_4T_{mlm}^6\,\rho N_{li}-\frac{18}{35}b_4T_{lli}^6\,\rho\nn\\
&-\frac{2}{7}b_4T_{lim}^6\,\rho N_{lm}.
\end{eqnarray}

\begin{eqnarray}\label{pol}
\dot{P}_i^S=&-\frac{4}{3}\pi\ell_c^3\,U_0\,\partial_j\,\Bigg(D_2\,\frac{\rho}{7}\Bigg(\frac{7}{15}\,\left[\partial_i(\rho P_j)+\partial_j(\rho P_i)+\delta_{ij}\boldsymbol\nabla\cdot(\rho\textbf{P})\right]\nn\\
&+\partial_k(\rho P_j)N_{ik}+\partial_j(\rho P_l)N_{li}+\partial_k(\rho P_i)N_{kj}+\boldsymbol\nabla\cdot(\rho\textbf{P})N_{ij}\nn\\
&+\partial_k(\rho P_l)N_{lk}\,\delta_{ij}+\partial_i(\rho P_l)N_{lj}\Bigg)+D_\perp \rho\left(\partial_j(\rho P_l)N_{li}+\frac{1}{3}\partial_j(\rho P_i)\right)\Bigg)\nn\\
&+\frac{4}{3}\pi D_R\ell_c^3 U_0\rho^2 \left(\frac{2}{3}P_i - P_jN_{ij}\right),
\end{eqnarray}

\begin{eqnarray}
\dot{N}_{ij}^L=&-\boldsymbol\nabla . \left[\left(b_1\textbf{T}^{1}+b_2\,\textbf{T}^{2} \right)\rho N_{ij}\right]\nn\\
&-\frac{1}{5}b_3\partial_z \left(T_{zj}^3\rho P_i+T_{zi}^3\rho P_j-\frac{2}{3}\delta_{ij}T_{zl}^3\rho P_l\right)\nn\\
&+\frac{\rho}{7}\Bigg(\frac{7}{5}\left[b_1(T_{ij}^4+T_{ji}^4)-b_2(T_{ij}^5+T_{ji}^5)\right]+5b_1T_{il}^4\,N_{jl}\nn\\
&-\frac{14}{15}b_1T_{ll}^4\,\delta_{ij}-2b_1T_{li}^4\,N_{lj}-3b_2T_{il}^5N_{lj}-2b_1T_{lj}^4N_{li}+5b_1T_{jl}^4N_{li}\nn\\
& -3b_2T_{lj}^5N_{li}-2b_1T_{ll}^4N_{ij}-2\left[b_1T_{ml}^4-b_2T_{ml}^5\right]N_{ml}\delta_{ij}\Bigg)\nn\\
&+\frac{4}{315}b_4\rho\Bigg(-9T_{lli}^6P_j+26T_{ijl}^6P_l-9T_{llj}^6P_i-9T_{llm}^6P_m\delta_{ij}\nn\\
&+26T_{jil}^6P_l-9T_{lij}^6P_l\Bigg).
\end{eqnarray}

\begin{eqnarray} \label{nem}
\dot{N}_{ij}^S=&-\frac{4}{3}\pi\ell_c^3\,U_0\,\partial_k\Bigg(\rho\Bigg(\frac{1}{35}D_2\Bigg(\partial_l(\rho P_l)P_i\,\delta_{jk}+\partial_l(\rho P_l)P_j\,\delta_{ik}\nn\\
&-\frac{4}{3}\,(\partial_l(\rho P_l)P_k+\partial_l(\rho P_k)P_l)\,\delta_{ij}+\partial_l(\rho P_i)P_l\,\delta_{jk}\nn\\
 &+\partial_l(\rho P_j)P_l\,\delta_{ik}+\partial_i(\rho P_m)P_m\,\delta_{jk}+\partial_j(\rho P_m)P_m\,\delta_{ik}\nn\\
&+(\partial_j(\rho P_i)+\partial_i(\rho P_j))P_k+\partial_j(\rho P_k)P_i+\partial_i(\rho P_k)P_j \Bigg)\nn\\
&-\frac{2}{21}D_5\,\partial_k(\rho P_l)P_l\,\delta_{ij}+\frac{1}{5}D_4\,\partial_k\left[(\rho P_i)P_j+(\rho P_j)P_i])\right]\Bigg)\Bigg)\nn\\
&+\frac{8}{5}\pi D_R\ell_c^3 U_0\rho^2\left(P_i P_j-\frac{P^2}{3}\,\delta_{ij}\right).
\end{eqnarray}

In sections \ref{stabilityanalysisI} and \ref{stabilityanalysisII} where, we  studied the linear stability of 
isotropic and polar phases we needed to linearize the interaction contributions.  Here we briefly 
present the details of such calculations for a typical term. Let us consider  the first term of 
$\dot{\rho}^L$ in density equation (\ref{hyd-rho}).  We have:
\begin{eqnarray}\label{lin}
-b_1\partial_i \left(T_i^1\rho\right)=&-b_1\partial_i \left(\int {\rm d}\textbf{r}^\prime\frac{\hat{R}_i}{R^2}M_{jk}(\hat{\textbf{R}})\rho(\textbf{r}^\prime)N_{jk}(\textbf{r}^\prime)\rho(\textbf{r})\right),
\end{eqnarray}
where $\textbf{R}=\textbf{r}-\textbf{r}^\prime$. Now we can substitute isotropic  values of $\rho$ and $\textbf{N}$ from (\ref{iso-rho}) and (\ref{iso-N}) to (\ref{lin}).  Taking the spatial Fourier transform and  
defining $W_{ijk}=\frac{\hat{R}_i}{R^2}M_{jk}$, we will have:
\begin{eqnarray}\label{FT}
-b_1\partial_i \left(T_i^1\rho\right)&=-b_1\rho_0^2\,\partial_i \left(\int {\rm d}\textbf{r}^\prime\int {\rm d}\textbf{k} \rme^{-\rmi\textbf{k}\cdot(\textbf{r}-\textbf{r}^\prime)}\check{W}_{ijk}(\textbf{k})\int {\rm d}\textbf{k}^\prime \rme^{-\rmi\textbf{k}^\prime\cdot\textbf{r}^\prime}\delta \check{N}_{jk}(\textbf{k}^\prime)\right)\nn\\
&=-b_1\rho_0^2\,\partial_i \left(\int {\rm d}\textbf{k}\int {\rm d}\textbf{k}^\prime \delta(\textbf{k}-\textbf{k}^\prime)\check{W}_{ijk}(\textbf{k})\rme^{-\rmi\textbf{k}\cdot\textbf{r}}\delta \check{N}_{jk}(\textbf{k}^\prime)\right)\nn\\
&=\rmi b_1\rho_0^2\left(\int {\rm d}\textbf{k}\rme^{-\rmi\textbf{k}\cdot\textbf{r}}\,k_i\,\check{W}_{ijk}(\textbf{k})\delta \check{N}_{jk}(\textbf{k})\right).
\end{eqnarray}
To proceed further, we need to calculate the Fourier transform of $W_{ijk}$. As a result of symmetry, the 
following general expression for  $\check{W}_{ijk}$ can be written:
\begin{eqnarray}
\check{W}_{ijk}=A\hat{k}_i\hat{k}_j\hat{k}_k+B\hat{k}_i\delta_{jk}+C\hat{k}_j\delta_{ik}+D\hat{k}_k\delta_{ij},
\end{eqnarray}
where scalar functions $A$, $B$, $C$ and $D$ can depend on $k$ and Fourier transform is defined by:
\begin{eqnarray}
\check{W}_{ijk}=\int {\rm d}\textbf{R}\,\rme^{\rmi\textbf{k}\cdot\textbf{R}}\,W_{ijk}(\textbf{R}).
\end{eqnarray}
Multiplying the above two equations by $\hat{k}_i\hat{k}_j\hat{k}_k$, $\hat{k}_i\delta_{jk}$, $\hat{k}_j\delta_{ik}$ and $\hat{k}_k\delta_{ij}$ respectively, we will 
obtain the following four equations for unknown functions:
\begin{eqnarray}
A+B+C+D=\int {\rm d}\textbf{R}\,\rme^{\rmi\textbf{k}\cdot\textbf{R}}\frac{1}{R^2}\left((\hat{\textbf{k}}\cdot\hat{\textbf{R}})^3-\frac{1}{3}(\hat{\textbf{k}}\cdot\hat{\textbf{R}})\right),
\end{eqnarray}
\begin{eqnarray}
A+3B+C+D=0,
\end{eqnarray}
\begin{eqnarray}
A+B+3C+D=\frac{2}{3}\int {\rm d}\textbf{R}\,\rme^{\rmi\textbf{k}\cdot\textbf{R}}\frac{1}{R^2}(\hat{\textbf{k}}\cdot\hat{\textbf{R}}),
\end{eqnarray}
\begin{eqnarray}
A+B+C+3D=\frac{2}{3}\int {\rm d}\textbf{R}\,\rme^{\rmi\textbf{k}\cdot\textbf{R}}\frac{1}{R^2}(\hat{\textbf{k}}\cdot\hat{\textbf{R}}).
\end{eqnarray}
By evaluating the integrals, we can obtain the following result for $ \check{W}_{ijk}$:
\begin{eqnarray}
 \check{W}_{ijk}=-\frac{8\pi\rmi}{3k}\hat{k}_i\hat{k}_j\hat{k}_k+\frac{4\pi\rmi}{3k}\hat{k}_j\delta_{ik}+\frac{4\pi\rmi}{3k}\hat{k}_k\delta_{ij}.
\end{eqnarray}
With a similar procedure, all other interaction integrals can be calculated.

\section*{References}


\begin{thebibliography}{10}

\bibitem{Ramaswamy}
J~Toner, Y~Tu, and S~Ramaswamy.
\newblock Hydrodynamics and phases of flocks.
\newblock {\em Annals of Physics}, 318:170--244, 2005.

\bibitem{Marchetti3}
M~C Marchetti, J~F Joanny, S~Ramaswamy, T~B Liverpool, J~Prost, M~Rao, and R~A
  Simha.
\newblock Hydrodynamics of soft active matter.
\newblock {\em Reviews of Modern Physics}, 85:1143, 2013.

\bibitem{Vicsek}
T~Vicsek and A~Zafeiris.
\newblock Collective motion.
\newblock {\em Physics Reports}, 517:71--140, 2012.

\bibitem{Ramaswamy5}
S~Ramaswamy.
\newblock Active matter.
\newblock {\em Journal of Statistical Mechanics: Theory and Experiment},
  2017:054002, 2017.

\bibitem{Becco}
Ch~Becco, N~Vandewalle, J~Delcourt, and P~Poncin.
\newblock Experimental evidences of a structural and dynamical transition in
  fish school.
\newblock {\em Physica A: Statistical Mechanics and its Applications},
  367:487--493, 2006.

\bibitem{Cavagna}
A~Cavagna, L~Del~Castello, I~Giardina, T~Grigera, A~Jelic, S~Melillo, T~Mora,
  L~Parisi, E~Silvestri, M~Viale, and A~M Walczak.
\newblock Flocking and turning: a new model for self-organized collective
  motion.
\newblock {\em Journal of Statistical Physics}, 158:601--627, 2015.

\bibitem{Nagy}
M~Nagy, Z~Akos, D~Biro, and T~Vicsek.
\newblock Hierarchical group dynamics in pigeon flocks.
\newblock {\em Nature}, 464:890--893, 2010.

\bibitem{Peruani}
F~Peruani, J~Starru\ss{}, V~Jakovljevic, L~S\o{}gaard-Andersen, A~Deutsch, and
  M~B{\"a}r.
\newblock Collective motion and nonequilibrium cluster formation in colonies of
  gliding bacteria.
\newblock {\em Physical Review Letters}, 108:098102, 2012.

\bibitem{Tokita}
R~Tokita, T~Katoh, Y~Maeda, J-i Wakita, M~Sano, T~Matsuyama, and M~Matsushita.
\newblock Pattern formation of bacterial colonies by escherichia coli.
\newblock {\em Journal of the Physical Society of Japan}, 78:074005, 2009.

\bibitem{Lushi}
E~Lushi, H~Wioland, and R~E Goldstein.
\newblock Fluid flows created by swimming bacteria drive self-organization in
  confined suspensions.
\newblock {\em Proceedings of the National Academy of Sciences},
  111:9733--9738, 2014.

\bibitem{Sokolov}
A~Sokolov, I~S Aranson, J~O Kessler, and R~E Goldstein.
\newblock Concentration dependence of the collective dynamics of swimming
  bacteria.
\newblock {\em Physical Review Letters}, 98:158102, 2007.

\bibitem{Prost}
J~Prost, F~J{\"u}licher, and JF~Joanny.
\newblock Active gel physics.
\newblock {\em Nature Physics}, 11:111--117, 2015.

\bibitem{Theurkauff}
I~Theurkauff, C~Cottin-Bizonne, J~Palacci, C~Ybert, and L~Bocquet.
\newblock Dynamic clustering in active colloidal suspensions with chemical
  signaling.
\newblock {\em Physical Review Letters}, 108:268303, 2012.

\bibitem{Speck}
I~Buttinoni, J~Bialk{\'e}, F~K{\"u}mmel, H~L{\"o}wen, C~Bechinger, and T~Speck.
\newblock Dynamical clustering and phase separation in suspensions of
  self-propelled colloidal particles.
\newblock {\em Physical Review Letters}, 110:238301, 2013.

\bibitem{Bayati}
P~Bayati and A~Najafi.
\newblock Dynamics of two interacting active janus particles.
\newblock {\em Journal of Chemical Physics}, 144:134901, 2016.

\bibitem{Narayan}
V~Narayan, S~Ramaswamy, and N~Menon.
\newblock Long-lived giant number fluctuations in a swarming granular nematic.
\newblock {\em Science}, 317:105--108, 2007.

\bibitem{Schaller}
V~Schaller, C~Weber, C~Semmrich, E~Frey, and A~R Bausch.
\newblock Polar patterns of driven filaments.
\newblock {\em Nature}, 467:73--77, 2010.

\bibitem{Sumino}
Y~Sumino, K~H Nagai, Y~Shitaka, D~Tanaka, K~Yoshikawa, H~Chat{\'e}, and K~Oiwa.
\newblock Large-scale vortex lattice emerging from collectively moving
  microtubules.
\newblock {\em Nature}, 483:448--452, 2012.

\bibitem{Mermin}
N~D Mermin and H~Wagner.
\newblock Absence of ferromagnetism or antiferromagnetism in one or
  two-dimensional isotropic heisenberg models.
\newblock {\em Physical review letters}, 17:1133, 1966.

\bibitem{Tu}
J~Toner and Y~Tu.
\newblock Long-range order in a two-dimensional dynamical xy model: How birds
  fly together.
\newblock {\em Physical Review Letters}, 75:4326, 1995.

\bibitem{Marchetti2}
A~Baskaran and M~C Marchetti.
\newblock Statistical mechanics and hydrodynamics of bacterial suspensions.
\newblock {\em Proceedings of the National Academy of Sciences}, 106:15567,
  2009.

\bibitem{Stark3}
J~Blaschke, M~Maurer, K~Menon, A~Z{\"o}ttl, and H~Stark.
\newblock Phase separation and coexistence of hydrodynamically interacting
  microswimmers.
\newblock {\em Soft matter}, 12:9821--9831, 2016.

\bibitem{Shelley}
D~Saintillan and M~J Shelley.
\newblock Instabilities and pattern formation in active particle suspensions:
  kinetic theory and continuum simulations.
\newblock {\em Physical Review Letters}, 100:178103, 2008.

\bibitem{Behmadi}
H~Behmadi, Z~Fazli, and A~Najafi.
\newblock A 2d suspension of active agents: the role of fluid mediated
  interaction.
\newblock {\em Journal of Physics: Condensed Matter}, 29:115102, 2017.

\bibitem{Shelley2}
D~Saintillan and M~J Shelley.
\newblock Instabilities, pattern formation, and mixing in active suspensions.
\newblock {\em Physics of Fluids}, 20:123304, 2008.

\bibitem{inst1}
C~A Whitfield, T~C Adhyapak, A~Tiribocchi, G~P Alexander, D~Marenduzzo, and
  S~Ramaswamy.
\newblock Hydrodynamic instabilities in active cholesteric liquid crystals.
\newblock {\em The European Physical Journal E}, 40:50, 2017.

\bibitem{inst2}
N~Oyama, J~J Molina, and R~Yamamoto.
\newblock Simulations of model microswimmers with fully resolved hydrodynamics.
\newblock {\em Journal of the Physical Society of Japan}, 86:101008, 2017.

\bibitem{inst3}
M~M Genkin, A~Sokolov, O~D Lavrentovich, and I~S Aranson.
\newblock Topological defects in a living nematic ensnare swimming bacteria.
\newblock {\em Physical Review X}, 7:011029, 2017.

\bibitem{Rao1}
S~Ramaswamy and M~Rao.
\newblock Active-ﬁlament hydrodynamics: instabilities, boundary conditions
  and rheology.
\newblock {\em New Journal of Physics}, 9:423, 2007.

\bibitem{Stark}
A~Z\"{o}ttl and H~Stark.
\newblock Hydrodynamics determines collective motion and phase behavior of
  active colloids in quasi-two-dimensional confinement.
\newblock {\em Physical Review Letters}, 112:118101, 2014.

\bibitem{Peruani2}
F~Peruani, A~Deutsch, and M~B{\"a}r.
\newblock Nonequilibrium clustering of self-propelled rods.
\newblock {\em Physical Review E}, 74:030904, 2006.

\bibitem{Vicsek2}
T~Vicsek, A~Czir{\'o}k, E~Ben-Jacob, I~Cohen, and O~Shochet.
\newblock Novel type of phase transition in a system of self-driven particles.
\newblock {\em Physical review letters}, 75:1226, 1995.

\bibitem{Peruani3}
H~Chat{\'e}, F~Ginelli, G~Gr{\'e}goire, F~Peruani, and F~Raynaud.
\newblock Modeling collective motion: variations on the vicsek model.
\newblock {\em The European Physical Journal B}, 64:451--456, 2008.

\bibitem{Marchetti5}
A~Baskaran and M~C Marchetti.
\newblock Enhanced diffusion and ordering of self-propelled rods.
\newblock {\em Physical Review Letters}, 101:268101, 2008.

\bibitem{Tu1}
J~Toner and Y~Tu.
\newblock Flocks, herds, and schools: A quantitative theory of flocking.
\newblock {\em Physical Review E}, 58:4828, 1998.

\bibitem{Ramaswamy4}
S~Ramaswamy and R~A Simha.
\newblock Hydrodynamic fluctuations and instabilities in ordered suspensions of
  self-propelled particles.
\newblock {\em Physical Review Letters}, 89:058101, 2002.

\bibitem{Mishra}
S~Mishra, A~Baskaran, and M~C Marchetti.
\newblock Fluctuations and pattern formation in self-propelled particles.
\newblock {\em Physical Review E}, 81:061916, 2010.

\bibitem{Bertin}
E~Bertin, M~Droz, and G~Gr{\'e}goire.
\newblock Hydrodynamic equations for self-propelled particles: microscopic
  derivation and stability analysis.
\newblock {\em Journal of Physics A: Mathematical and Theoretical}, 42:445001,
  2009.

\bibitem{Purcell}
E~M Purcell.
\newblock Life at low reynolds number.
\newblock {\em American Journal of Physics}, 45:3--11, 1977.

\bibitem{3SJPC}
A~Najafi and R~Golestanian.
\newblock Propulsion at low reynolds number.
\newblock {\em Journal of Physics: Condensed Matter}, 17:S1203, 2005.

\bibitem{3Sfaez}
R~Zargar, A~Najafi, and MF~Miri.
\newblock Three-sphere low-reynolds-number swimmer near a wall.
\newblock {\em Physical Review E}, 80:026308, 2009.

\bibitem{Pozrikidis}
C~Pozrikidis.
\newblock {\em Boundary integral and singularity methods for linearized viscous
  flow}.
\newblock Cambridge University Press, 1992.

\bibitem{Moradi}
M~Moradi and A~Najafi.
\newblock Rheological properties of a dilute suspension of self-propelled
  particles.
\newblock {\em Europhysics Letters}, 109:24001, 2015.

\bibitem{Farzin}
M~Farzin, K~Ronasi, and A~Najafi.
\newblock General aspects of hydrodynamic interactions between three-sphere low
  reynolds number swimmers.
\newblock {\em Physical Review E}, 85:061914, 2012.

\bibitem{Yeomans4}
G~P Alexander, C~M Pooley, and J~M Yeomans.
\newblock Hydrodynamics of linked sphere model swimmers.
\newblock {\em Journal of Physics: Condensed Matter}, 21:204108, 2009.

\bibitem{Lauga}
E~Lauga and T~R Powers.
\newblock The hydrodynamics of swimming microorganisms.
\newblock {\em Reports on Progress in Physics}, 72:096601, 2009.

\bibitem{Elgeti}
J~Elgeti, R~G Winkler, and G~Gompper.
\newblock Physics of microswimmers—single particle motion and collective
  behavior: a review.
\newblock {\em Reports on Progress in Physics}, 78:056601, 2015.

\bibitem{Drescher}
K~Drescher, R~E Goldstein, N~Michel, M~Polin, and I~Tuval.
\newblock Direct measurement of the flow field around swimming microorganisms.
\newblock {\em Physical Review Letters}, 105:168101, 2010.

\bibitem{coherentcoupling}
A~Najafi and R~Golestanian.
\newblock Coherent hydrodynamic coupling for stochastic swimmers.
\newblock {\em Europhysics Letters}, 90:68003, 2010.

\bibitem{Kim}
S~Kim and S~J Karrila.
\newblock {\em Microhydrodynamics Principles and Selected Applications}.
\newblock Dover Publications, Inc., New York, 2005.

\bibitem{Happel}
J~Happel and H~Brenner.
\newblock {\em Low Reynolds Number Hydrodynamics with special applications to
  particulate media}.
\newblock Prentice-Hall, Englewood Cliffs, NJ, 1965.

\bibitem{Liverpoolsquiremer}
N~Yoshinaga and T~B Liverpool.
\newblock Hydrodynamic interactions in dense active suspensions: From polar
  order to dynamical clusters.
\newblock {\em Physical Review E}, 96:020603, 2017.

\bibitem{Howse}
J~R Howse, R~A~L Jones, A~J Ryan, T~Gough, R~Vafabakhsh, and R~Golestanian.
\newblock Self-motile colloidal particles: From directed propulsion to random
  walk.
\newblock {\em Physical Review Letters}, 99:048102, 2007.

\bibitem{Marchetti}
A~Baskaran and M~C Marchetti.
\newblock Hydrodynamics of self-propelled hard rods.
\newblock {\em Physical Review E}, 77(6):011920, 2008.

\end{thebibliography}

\end{document}